# Social Diversity and Spread of Pandemic: Evidence from India

Upasak Das

Global Development Institute, University of Manchester

(upasak.das@manchester.ac.uk)

Udayan Rathore

Oxford Policy Management

(rathore.udayan@gmail.com)

Prasenjit Sarkhel

Kalyani University

(prasenjitsarkhel@gmail.com)

## Abstract

*Existing studies have documented how social diversity can influence public good provisioning, economic development and other welfare outcomes. However, evidence on the implications of social diversity during pandemics is limited. Compliance with the public health guidelines requires coordinated community actions which might be undermined in diverse areas. In this paper, we assess the relationship between caste-group diversity and the spread of COVID-19 infection during the nationwide lockdown and unlocking period in India. On the extensive margin, we find that caste-homogeneous districts systematically took more days to cross the concentration thresholds of 50 to 500 cases. Estimates on the intensive margin, using daily cases, further show that caste-homogeneous districts experienced slower growth in infection. Overall, the effects of caste-group homogeneity remained positive and statistically significant for 2.5 months (about 76 days) after the beginning of the lockdown and weakened with subsequent phases of the lockdown. The results hold even after accounting for the emergence of initial hotspots before lockdown, broader diffusion patterns through daily fixed effects, region fixed effects, and dynamic administrative response through time-variant lagged COVID-19 fatalities at the district level. These effects are not found to be confounded by differential levels of testing and underreporting of cases in some states. Consistent estimates from bias-adjusted treatment effects also ensure that our findings remain robust even after accounting for other unobservables. We find suggestive evidence of higher engagement of community health workers in caste-homogenous localities, which further increased after the lockdown. We posit this as one potential channel that can explain our results. Our findings reveal how caste-group diversity can be used to identify potential hotspots during public health emergencies and emphasize the importance of community health workers and decentralized policy response.*

## 1. Introduction

Ethnic and social group affiliations often shape individual and group identities. These in turn influence cooperation and coordination within and across communities, which is particularly paramount during times of crises. For example, such affiliations serve as an important channel for information transmission (Larson & Lewis, 2017) and are likely to ensure compliance with social norms (Goette, Huffman & Meir, 2006). In this paper, we ask if community-level diversity (or homogeneity) can influence the overall public health outcomes during pandemics? Specifically, we study how caste-group diversity affected the spread of the SARS-CoV-2 viral infection (henceforth COVID-19) during the early phase of the pandemic in India. Here, we first assess if the time taken to cross specific thresholds of COVID-19 cases increases in districts with higher levels of caste-group diversity. We also examine the implications of caste-group diversity on the progression of daily cases during the nationwide lockdown and an equivalent period of unlocking .

During the first wave of infection in early 2020, compliance with the appropriate COVID-19 norms was widely recommended as the only protective measure against infection.[1] Ensuring compliance with these norms requires coordinated community action, more so because private protective effort involves a positive externality. Unless other community members follow compliance protocols, the effectiveness of an individual's effort is negligible. However, a large body of evidence from developing countries suggests diversity can itself undermine voluntary contribution to public goods (Banerjee et al., 2005; Habyarimana et al., 2007). If preference is biased towards in-group members than outsiders (Alesina and Ferrara, 2005) or more diverse societies lack social capital by virtue of lower social interactions and interpersonal trust, a community's ability to effectively monitor and sanction violations of norms might be restricted, thereby weakening collective action (Banerjee et al., 2005; Miguel and Gugerty, 2005)[2]. These channels suggest that diverse communities may experience a faster spread of COVID-19 infection.

---

[1] India commenced its vaccination drive only on 16th January 2021.

[2], Greater diversity is also associated with divergence in preferences and varied risk perception that may lead to a sub-optimal level of compliance (Alesina et al., 1999).

However, some recent studies have challenged this conventional wisdom and show that diversity, under certain conditions, can also support welfare gains (Gisselquist et al., 2016). In the context of the pandemic in Russia, Egorov et al. (2021) argue that infected individuals in diverse communities are less altruistic and hence have a lower incentive to self-isolate. This knowledge forces the uninfected population to reduce mobility, thereby arresting the spread of the virus. In addition, infection growth in diverse communities may also be inhibited by lower community interactions (Bosancianu, et al. 2020). These countervailing possibilities turn the potential relationship between social diversity and infection spread into an ambiguous one that necessitates empirical investigation

The question assumes importance in the context of India as it is also one of the most socially diverse countries in the world that is stratified on the basis of caste and caste groups (Deshpande 2000; Munshi, 2019). Not only is caste a dominant force in India's social structure, but it is also an essential basis of social network formation (Thorat, 2010; Desai and Dubey, 2011; Munshi, 2019). In this paper, we consider all the four prominent caste groups in India. These include the Upper Caste (Others), the socially deprived Scheduled Caste (SC, formerly untouchable caste), the ethnic and tribal groups belonging to Scheduled Tribes (ST) and the Other Backward Castes (OBC). Literature on residential patterns in rural India, where much of the population resides, shows that households follow distinct spatial segregation whereby members of a common caste live in a particular area (Munshi, 2019). These clusters are associated with high interpersonal trust, cooperation and compliance, which may impede infection transmission at the community level.

During the initial phase of the pandemic, and like other countries, India also relied on mobility restrictions through nationwide lockdowns. These were imposed from March 25 to May 31, 2020, in four continuous phases to minimize mobility and control the spread of the virus, with minor relaxations introduced over time.[3] These mobility restrictions at the time were seen as one of the strictest ones imposed anywhere in the world (Gisselquist and Kundu 2020). Thereafter, from June 1, 2020, comprehensive unlocking measures were initiated, and respective state governments were given more flexibility to decide on restrictions outside a countrywide negative list that included international travel and the opening of educational institutions.[4]

[3] https://www.bbc.com/news/world-asia-india-52290761 (accessed on December 12, 2021).

In this paper, we first explore the relationship of interest on the extensive margin and check if caste-group diversity exerted any influence on the average number of days a district took to cross a certain threshold of COVID-19 cases.[5] Here, we use multiple threshold values ranging from 50 to 500 cases.[6] To track the infection trajectory over time on the intensive margin, we examine if there was a significant differential in the spread of COVID-19 cases as shares of its population in caste-group homogeneous districts using daily COVID-19 data. In particular, we ask if caste-diverse districts in India experienced faster infection spread during the lockdown and whether these effects, if any, persisted even after these restrictions were withdrawn. Here, we consider the entire lockdown period that lasted for 68 days (March 25 to May 31, 2020) and an equivalent number of days during the unlocking period (June 1 to August 7, 2020). It is important to note that India recorded more than 2 million cases and close to 44,000 COVID-19 related deaths in this period.

Our identification strategy is based on controlling for a wide range of confounding possibilities to isolate the relationship of interest (see Section 2). In addition to a range of socio-economic, demographic, and movement & migration-related variables, we also control for any COVID-19 cases in the district before the nationwide lockdown and time-invariant region-specific fixed effects. Importantly, in the day-wise analysis, we hypothesize that the administrative machinery and public health authorities are likely to be more sensitive to recent deaths when allocating scarce resources. Accordingly, we control for the rolling number of COVID-19 deaths as a share of the population in the past seven days as a measure of time-variant supply-side response. We also undertake a range of internal validity tests to verify our findings and account for alternate channels of infection transmission (Section 3.3). Importantly, we account for the influence of unobserved heterogeneity that provide consistent estimates of bias-adjusted treatment effects to further support our findings (Altonji et al. 2005; Oster, 2019).

[4] https://timesofindia.indiatimes.com/india/lockdown-4-0-after-may-31-states-to-have-own-curbs-lists/articleshow/76024524.cms (accessed on December 12, 2021)

[5] We consider days taken to record 50, 100, 200, 300, 400 and 500 cases as the respective thresholds for the period between March 25, 2020 to March 21, 2021.

[6] In this analysis, we only consider the districts which crossed the specified threshold in question. Although our interest is in the relationship between caste-group homogeneity and spread of infection during the nationwide lockdown and a comparable period of unlocking, taking this entire period for which, the data is available ensures that districts, especially for larger thresholds, do not drop out of the analysis.

We find that caste-homogeneous districts, on average, took significantly more days to breach the thresholds of 50 to 500 COVID-19 cases. Lower absolute number of cases in these regions may have provided more time to public health authorities to mount an appropriate policy response. Our day-wise results indicate significantly slower growth in COVID-19 cases in areas characterized by higher caste-group homogeneity. These gains dampen over time but remain significant throughout the lockdown period, lasting for about a week after unlocking. The results are additionally robust to potential biases emerging from factors such as differences in testing rates across states, alternate measures of caste-group diversity and disproportionately lower reporting of cases in some states.

We further offer suggestive mechanisms that can explain our findings. First, we make use of the fact that community health workers (CHW), who are the main point of contact with the primary health care system, were mobilized as frontline workers during COVID-19. Their task included engaging with the community to generate awareness about COVID-19 along with case identification, contact tracing and reporting potential infections to district health teams (Salve et al., 2023). These CHWs are generally recruited from the same community, which they serve. Hence, their effectiveness in performing the aforementioned tasks can potentially be higher in socially homogenous communities, which are characterised by higher reciprocity and trust that can foster higher cooperation (Gadsden et.al. 2022). Our observations using data from the National Family Health Survey dataset indicate disproportionately higher involvement of CWHs in caste-group homogeneous areas after the first COVID-19 lockdown. We argue that this higher involvement was potentially influential in reducing the spread of the infection. Second, using the Time-Use Survey conducted in 2019, we find higher social interaction and involvement in unpaid voluntary social activities among the residents in caste-group homogeneous areas. The higher levels of socialization and participation in social activities may facilitate higher social capital formation and strengthen collective action, resulting in slower infection spread in caste homogeneous regions.

Our main contributions to the literature are as follows. One, we contribute to the evidence on social diversity and health outcomes in a pandemic setting and find that spread of COVID-19 was faster in caste-group diverse districts. In doing so, we also contribute to the growing literature examining the implications of social diversity and provisioning of public goods tied to public health emergencies. Chakraborty and Mukherjee (2023) looked at the relationship of economic inequality on heterogeneity of COVID-19 spread in India. In this paper, we control for economic inequality and then explored the influence of caste diversity. Two, from a policy perspective, our findings reveal how caste-group diversity can be used to identify potential hotspots during pandemics. We argue that prioritizing the allocation of scarce medical and administrative resources to these vulnerable areas can slow overall infection growth. These policy prescriptions remain pertinent in the wake of the new strains of concern and inform future public health crises. Third, our research indicates that health interventions through CHWs might be relatively more effective during health emergencies. Taking this aspect into cognizance might require a mix of centralized and decentralized health interventions for formulating effective health policies that would also ensure their involvement in socially diverse areas for future pandemics.

The paper is structured as follows. Section 2 describes the data used and discusses the variables used and the identification strategy. Section 3 provides the regression results with robustness checks and explores the possible mechanism that explains our findings. Section 4 discusses the policy implications, and section 5 concludes.

## 2. Data, variables and identification strategy

### *2.1 Data*

District-day-level data on COVID-19 cases and deaths come from the Development Data Lab's (DDL) COVID India database, available for 684 of 718 (95%) districts in India, which is the unit of analysis.[7] For analysis on the extensive margin, we estimate the number of days taken to cross thresholds from 50 to 500 at the district level, counted from the beginning of the lockdown.[8] The analysis spans until March 21, 2021, and our research does not consider any district that crossed the relevant threshold after this period. Therefore, if a district crossed 50 cases on March 28, 2020, the number of days in our case would be 4 (March 25 to March 28). The duration of analysis for the analysis on intensive margin ranges from March 25 to August 7, 2020. This covers two distinct periods: (a) the first nationwide lockdown, which ended on the May 31, 2020 and, (b) the unlocking phase of an equivalent duration (68 days) starting from June 1 to August 7, 2020.

---

[7] The Socioeconomic High-resolution Rural-Urban Geographic Platform for India (SHRUG) is an open source platform to facilitate data sharing between researchers working on India (http://www.devdatalab.org/shrug). These numbers are taken from covidindia.com, a crowd sourced project.

[8] The district level cases are available from beginning of February 2020 but no district recorded a total of 50 cases before lockdown (March 25, 2020).

We restrict our analysis to these 68 days of the lockdown and the equivalent period of unlocking for multiple reasons. First, even though the second wave of COVID-19 in 2021 was more severe, India recorded more than 2 million cases and close to 44,000 COVID-19 related deaths in the first wave during the period of consideration. Thus, the first wave of infection is important to analyze as it set the wheels in motion for subsequent spread and mutations of the virus that devastated India in 2021. Second, the Union (Central) Government of India imposed, and later eased, mobility restrictions during the lockdown based on a common set of criteria across all Indian states. These mobility restrictions at the time were seen as one of the strictest ones imposed anywhere in the world (Gisselquist and Kundu 2020). In contrast, during the second wave of infection in 2021, state governments imposed vastly differentiated, state-specific lockdowns and interventions based on localized surges in infections, fatalities, and medical preparedness. Restricting the analysis to the first nationwide lockdown and corresponding unlocking phase then allows us to minimize various state-specific unobserved influences.[9] Third, it can also be argued that knowledge about the spread of the virus during the first wave was limited, and the potential effects of not abiding by the COVID-19 appropriate protocols were largely considered detrimental, something that may not hold for later time periods (Brown et al., 2021). Notably, in the months right before the second wave of infection, a pre-mature victory over the pandemic was declared, which may have induced a varied and a false sense of protection from the virus and thereby may confound with the relationship of interest.[10]

[9] For the subsequent waves of infection, ascertaining the relationship of interest is further complicated by factors such as emergence of new strains that were not uniformly active across all Indian districts, events like the local body (Uttar Pradesh) and state elections that witnessed crowded political campaigns (Bihar, West Bengal, Tamil Nadu, Kerala, Puducherry and Assam), and the nationwide protests on the three farm laws whose intensity varied across different states, among others.

[10] https://timesofindia.indiatimes.com/education/news/pm-modi-cherishes-indias-dual-victory-over-covid-19-and-australia-praise-young-india/articleshow/80402121.cms (accessed on March 21, 2023)

We use multiple district-level datasets to obtain the set of explanatory variables (discussed in section 2.2.2). Data on the caste-group composition comes from the National Family Health Survey (NFHS-4) with appropriate sampling weights, which is mapped to daily district-level COVID-19 data.[11] The NFHS-4 survey conducted in 2015-16 collected information on 601,509 households across 640 districts at the time and is one of the most extensive and district-level representative household surveys in India.[12] Accordingly, relevant district representative measures for a range of socio-economic, demographic, amenities, health behavior and movement & migration-related variables are taken from this survey. Information on supply-side health indicators like the number of doctors and hospitals as a share of district-level population is taken from the latest Census round conducted in 2011. Average distances of the districts from state capitals and number of flights entering these districts for the duration of analysis are taken from Google Earth Pro and Airport Authority of India, respectively.

To understand the potential pathways behind the relationship of interest, we first use the household level data from the NFHS-5 (2019-2021) survey. Here, we utilize information on household's involvement with community health workers. In addition, to assess time allocation towards social interaction and unpaid voluntary community work, we use the Time Use Survey (TUS) conducted by the National Sample Survey Organization in 2019. This survey uses the International Classification of Activities for Time-Use Statistics, 2016, to provide information on the time duration of each activity performed by the sampled household members six years and above on the day before the survey.

### *2.2 Variables*

#### *2.2.1 Dependent Variables*

We first check for the relationship between the number of days taken by different districts to cross the threshold of 50 cases and caste-group homogeneity. We also explore the same using the threshold of

[11] We are able to map caste-group homogeneity and other controls for 590 of 684 districts for which COVID-19 data is available (86%).

[12] It is the Demographic and Health Survey (DHS) equivalent of India.

100, 200, 300, 400 and 500 cases. Then, using high-frequency daily data, we study the relationship between caste-group homogeneity and the spread of COVID-19 cases on the intensive margin. Here, we use the number of daily cases as a share of the population for each of the 136 days, starting from March 25 till August 7, 2020, as the primary outcome of interest.[13]

### *2.2.2 Explanatory variables*

Our main variable of interest is caste-group homogeneity, measured by the Herfindahl-Hirschman (HHI)/ fractionalization index at the district level. This index measures the probability that any two individuals chosen randomly from the residential unit belong to two different social groups. The HHI has previously been used extensively to measure social diversity or fragmentation in different contexts (Andreoni et al. 2016; Churchill et al. 2019). As mentioned previously, we use caste-groups that include the historically disadvantaged communities categorized under SCs, STs and OBCs alongside the privileged castes aggregated under "Others" to create this index.[14] Individuals from the ST community in India belong to particular social groups that maintain their traditions, norms and belief system, which are bounded by the religious, social and blood ties (Majumdar, 1958). Individuals from the SC community belong to the traditional Hindu society and have historically been part of the social group disparagingly referred to as *untouchables*. The Constitution of India (Article numbers 341 and 342) has recognized the SCs and STs as socially deprived groups. Those from the OBC community are also socially and economically disadvantaged but relatively better off than the STs and SCs. Individuals from upper castes have historically fared better on social, economic or educational outcomes as against other caste-groups, on average (Maity, 2017). The HHI or fractionalization index is given by $\sum_{n=1}^{4} s_n^2$ where $s_n$ is the population share of $n^{th}$ caste-group as described above. Figure-1 gives a spatial overview of the caste-group homogeneity across the Indian districts.

**Figure-1.**

[13] Appendix figure-1 shows the evolution in the cumulative number of cases across the districts over the four lockdown phases.

[14] The social networks within the Indian caste system are splintered into various sub-caste or jatis. In this paper, we restrict ourselves to the broad caste groupings that is representative at the district level.

Caste-group homogeneity across Indian districts

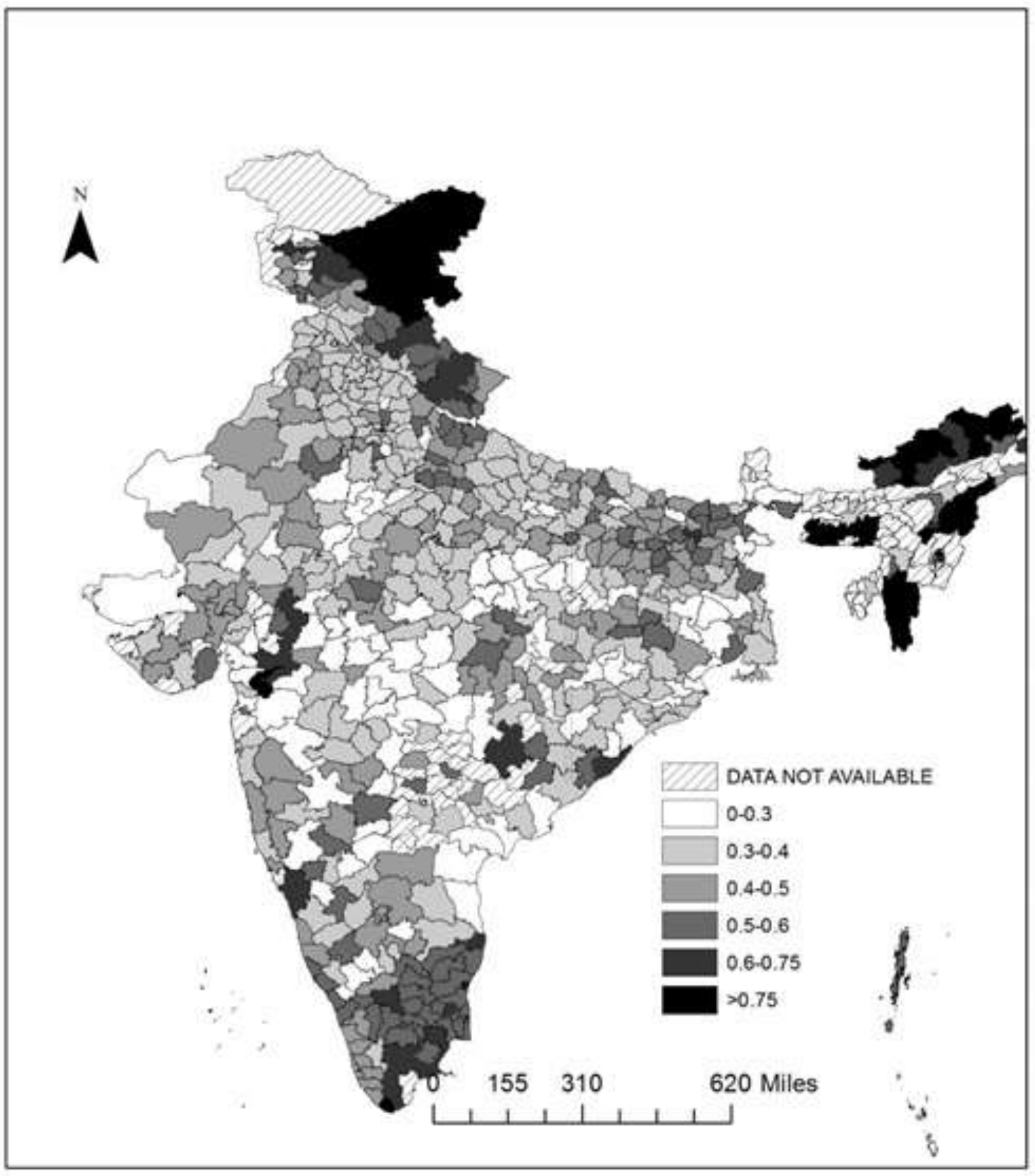

Note: ArcGIS software used to generate the figure. Caste-group homogeneity across Indian districts is measured using Herfindahl Hirschman Index. NFHS-4 dataset is used.

We use multiple control variables in the regression framework to account for confounding possibilities, as discussed in Section 2.3. As the emergence of initial hotspots is potentially endogenous, we also control for any cases reported in the district before the lockdown.[15] This is critical to address the selection on unobservables as it subsumes the influence of superior international connectivity and integration of certain districts into the global supply chain, among other factors. These are likely to be associated with residential patterns that shape diversity measures and infection spread at the district level. To account for the dynamic supply-side administrative responses in the day-wise analysis, we control for the rolling number of COVID-19 deaths as a share of the population in the past seven days in the district. Notably, this also potentially captures time-varying district-level unobservables that might confound our estimates. Public health in India is within the ambit of legislative and executive powers of the state and not the federal government. Here, each district is placed under the stewardship of a bureaucrat (district collector) that reports directly to the Chief Minister of a state. In such an administrative structure, the allocation of scarce administrative resources is likely to be sensitive to recent COVID-19 related casualties.

[15] The dummy variable takes a value of 1 if any cases were reported in the district on or before 24th March 2020 and 0 otherwise. Out of 590 districts used for analysis, 104 (17.6%) reported a total of 536 cases in this period. In an alternate specification and robustness check, we replace this dummy with the cases as a share of population before the lockdown.

Additionally, we control for a range of socio-economic factors in the dataset at the district level that includes shares of households that are: Hindus, have at least one member with an education level above senior secondary level, live in concrete houses and owner-owned dwellings as well as the share of households where members are covered by health insurance. We also control for the average numbers of rooms available for sleeping per household. This indicates the ability of household members to isolate in case of infection. To account for average district-level economic well-being, we use a standardized wealth index and the Gini coefficient that measures wealth inequality. Notably, studies have looked into the effects of economic inequality on COVID-19 spread (Chakraborty and Mukherjee, 2023) and here we account for the inequality to assess the implications of caste-group homogeneity. To control for access to public service announcements, share of households where a woman reads newspapers, listens to radio or watches television regularly is considered. For demographic profile and age-specific vulnerabilities to infection, we control for the share of households in a district with any elderly members (>=60 years of age), the density of population as well as the number of households in the district. To proxy for hygiene and health habits, we use share of households where soap was found in the handwashing area and where some member smokes at least on a weekly basis.

Moreover, we account for access to a range of amenities, which is likely to influence the economic resilience to the COVID-19 shock and the ability of the households to follow COVID-appropriate behavior. This includes share of households with access to improved drinking water, piped water inside the dwelling, toilets inside the housing compound, and clean cooking fuel. Also, we account for share of households that have access to social assistance in the form of below poverty line (BPL) cards, unique identification (Aadhar) documents and registered bank accounts, which may be used to identify the vulnerable and provide direct assistance. To control for dependence on public health facilities, we include share of households that usually go to public health centers when someone is unwell as a covariate. On the supply side, hospitals and doctors, as a share of the district-level population, are controlled for to account for physical and human health infrastructure. Although earlier studies in the pre-COVID-19 times have documented low levels of inter-district and inter-state migration in India (Munshi & Rosenzweig, 2009; Munshi, 2019), mobility and migration are likely to be key transmission channels. We control for short- and long-term migration from these districts to account for this possibility.[16] Moreover, we also account for long-term residential patterns of households in the district and the average distances of these districts from their state capital. The total number of flights entering the districts during the analysis period is also controlled for. Finally, in all the regressions, we take intra-state, administrative region fixed effects to account for the time-invariant, region-level unobserved characteristics. A detailed description of all variables used in the analysis and their summary statistics are provided in Appendix Table-1 and Appendix Table-2, respectively.

### *2.3 Empirical Strategy*

First, to study the relationship between days taken to record specific threshold of cases and caste-group homogeneity, we use the following regression specification:

$$NumDays_{dR} = \beta_0 + \beta_1 YDum_{dR}^0 + \beta_2 HISG_{dR} + \beta_3 C_{dR} + \pi_R + \epsilon_{dR} \qquad (1)$$

Here, $NumDays_d$ is the number of days taken to record a specific number of cases starting from March 25, 2020, in district, $d$ in region, $R$. The thresholds considered are 50, 100, 200, 300, 400 and 500 cases, respectively till March 21, 2021. So, if a district does not reach these thresholds by this date, it is not considered in our analysis. $YDum_{dR}^0$ is a dummy variable for any cases recorded before the nationwide lockdown that can account for initial hotspots. $HISG_{dR}$ represents standardized caste-group homogeneity measured through the HHI or fractionalization index. We use the standardized measure to ensure comparability across districts. The vector of time-invariant control variables at the district level, as discussed in Section 2.2.2, is captured by $C_{dR}$ and $\epsilon_{\mathrm{dR}}$ is the error term. Important to here is that the National Sample Survey Office (NSSO) classifies clusters of districts within a state under common NSSO regions (Government of India, 2001). These regions consist of contiguous districts with common geographical features, rural population density and cropping patterns, which may also influence the disease environment and spread of infection. Therefore, we account for the time-invariant region-level observed and unobserved characteristics through regional fixed effects given by $\pi_R$. Here, estimation is done using the standard Ordinary Least Squares (OLS) method with robust standard errors and $\beta_2$ represents the coefficient of interest. For the estimates on the intensive margin, we use the specification as in Lee et al. (2021) which examines the evolving relationship of interest on a daily basis. The regression equation is given by:

$$Y_{dRt} = \beta_0 + \beta_1 YDum_{dR}^{0} + \beta_2\, D_{DRt(lag7day)} + \sum_{t=1}^{136} \sigma_t \left(HISG_{dR} * Day_t\right) + \beta_4 C_{dR} + Day_t + +\pi_R + \epsilon_{dt} \qquad (2)$$

[16] India witnessed reverse migration from bigger cities during the nationwide lockdown and unlocking period. Controlling for mobility measures before the crisis would control for the potential exposure from reverse migration.

Here, $Y_{dRt}$represents the total number of cases as a share of population (per 10,000) for district $d$ from region, $R$ on day $t$. $YDum_{dR}^{0}$ is as specified in equation 1 and accounts for initial hotspots before the lockdown.[17] $D_{DRt(lag7day)}$ represents total deaths per 10,000 population reported in the seven days before day, $t$.[18] As discussed previously, $HISG_{dR}$ represents standardized caste-group homogeneity measured through the HHI or fractionalization index. $Day_t$ takes the value of 1 when the day is $t$ and 0 otherwise. Note that $t$ takes the value from 1 (for March 25, 2020, the first day of the lockdown) to 136 (for August 7, the last day of our analysis). $\sigma_t$ is the vector of coefficients of interest. The vector of time-invariant control variables at the district level and the region fixed effects, as discussed previously, are captured by $C_d$ and $\pi_R$ respectively. $Day_t$ is the vector of day fixed effects, which

[17] For all our estimates, if we replace this dummy with the number of pre-lockdown Covid-19 cases in the district per 10,000 population the results remain statistically comparable.
[18] In our robustness analysis, we changed this to a three-day lag and find similar results.

accounts for broader diffusion of COVID-19 transmission and trends in national administrative policies. $\epsilon_{dt}$ is the error term and standard errors are clustered two-way at day and district level.[19] [20]

Region-level fixed effects, emergence of initial hotspots before lockdown, lagged time-varying death count, and other extensive covariates potentially allow us to get closer to the true causal effects. Moreover, we carry out further checks to establish credibility of the inferences. To ensure that the unobserved heterogeneities do not confound our findings, we examine the potential variation in the estimates after accounting for unobservables through bias-adjusted treatment effect estimates that assumes selection on observables are proportional to selection on unobservables by a hypothetical parameter δ (Altonji et al. 2005; Oster, 2019). The method uses another hypothetical parameter ($R^2_{max}$), which comes from explained variation in the scenario where all potential unobserved variables are incorporated into the linear regression. While there can be various options to compute the bound on $R^2_{max}$, Oster (2019) proposes $R^2_{max} = 1.3 * R^2_0$ ($R^2_0$ is the $R$-squared value of the full model with observed control variables) using data from a comprehensive sample of randomized experiments.[21] The literature on this method assumes the effect of observables to be at least as important as unobservables and thus considers a bound on $\delta$ that lies between [-1,1] (Altonji et al. 2005, Rathore and Das 2021). The argument is as follows: if $|\delta|$ goes beyond 1, the variation in the outcome explained by unobservable(s) has to be higher than the combined influence of all the observables put together to push the effect size of the coefficient of the variable of interest to zero. With a comprehensive set of controls, this is likely to be improbable. Accordingly, using $R^2_{max}$, we estimate the value of $\delta$ for which the coefficient of interest turns zero. In addition, we also conduct a range of other robustness checks that are discussed in section 3.3.

## 3. Results

[19] In STATA 16, *reghdfe* command is used with *noabsorb* option for this.

[20] For robustness check, we add state-level day-wise testing data to the list of controls. See Results section for details. As testing figures are not consistently available for all states across the duration of the analysis, these are used for validation and not used in main estimation equation (2). The results remain valid if standard errors clustered two-way at day and state level.

[21] According to Oster (2019), the bounding value of the cut off at 1.3 allows effects from at-least 90% of the random experiments to survive.

### *3.1 Extensive Margin*

First, we check if caste-group homogeneous districts, controlling for other confounding possibilities, took more days to cross specific infection thresholds. Here, we first present the regression estimates for the average number of days taken by a district to report 50 cases, as outlined in equation (1). The regression results are presented in Table-1 where we introduce different layers of controls in six phases. Model 1 only controls for any pre-lockdown cases and district-level socio-economic variables. Model 2 adds migration, residential pattern, and mobility controls to the specification. In Model 3, we include covariates for demographic and health behavior controls. Models 4 and 5 also include amenities and physical and human health infrastructure, respectively. Model 6 represents the full specification where administrative region-specific time-invariant characteristics are additionally controlled. The findings from these set of regression estimates (see Table-1) show that a district with one standard deviation increase in standardized caste group homogeneity took about 4.5 to 16 additional days to record a total of 50 COVID-19 cases. Notably, these results remain positive and statistically significant as the control variables are introduced phase-wise. Thus, after accounting for multiple confounding influences, we find that caste-group homogeneous districts took a significantly greater number of days to record a total of 50 cases. In the context of the contagious nature of the virus, this would then suggest a distinct advantage in favor of caste-group homogeneous districts.

Table-1: Estimation for days taken to record 50 cases on caste-group homogeneity

| | (1) | (2) | (3) | (4) | (5) | (6) |
|---|---|---|---|---|---|---|
| | Model 1 | Model 2 | model 3 | Model 4 | Model 5 | Model 6 |
| Standardized HHISG | 14.5719*** | 16.3619*** | 7.7133*** | 9.0350*** | 9.1502*** | 4.4665** |
| | (2.2829) | (2.2984) | (2.3584) | (2.4389) | (2.5263) | (2.0536) |
| ***Controls for*** | | | | | | |
| -Dummy for any pre-lockdown cases | ✓ | ✓ | ✓ | ✓ | ✓ | ✓ |
| -Socio-economic variables | ✓ | ✓ | ✓ | ✓ | ✓ | ✓ |
| -Migration, Residential pattern and Mobility | × | ✓ | ✓ | ✓ | ✓ | ✓ |
| - Demography, Health behaviour | × | × | ✓ | ✓ | ✓ | ✓ |
| - Amenities | × | × | × | ✓ | ✓ | ✓ |
| -Human and Physical Health Infrastructure | × | × | × | × | ✓ | ✓ |
| -Administrative Region | × | × | × | × | × | ✓ |

| Fixed Effects | | | | | | |
|---|---|---|---|---|---|---|
| Constant | 71.4973*** | 128.8278*** | 338.6619*** | 331.0897*** | 335.8885*** | 241.1084*** |
| | (1.5242) | (26.4537) | (32.7956) | (36.7681) | (37.5932) | (39.0651) |
| Observations | 557 | 557 | 557 | 557 | 557 | 557 |
| $R^2$ | 0.357 | 0.379 | 0.484 | 0.503 | 0.503 | 0.820 |

Notes: OLS estimates of the effects of 1 SD increase in caste homogeneity are presented along with standard errors clustered two-way at day and district level in the parenthesis. Standardized HHI SG refers to the Z-score of Herfindahl Hirschman Index for caste-group homogeneity. *** p-value < 0.01, ** p-value < 0.05, * p-value <0.10.

We also verify this finding using a range of different thresholds of total COVID-19 cases. Using the same specification as in equation 1, we find a positive and statistically significant relationship for other thresholds ranging from 100 cases to 500 cases (see Figure-2). All these results point to initial infection concentration taking a relatively higher number of days in more caste-homogeneous regions. A natural follow-up question emerges on whether the overall spread of infection was slower in caste-group homogeneous districts, especially during the nationwide lockdown and whether these gained also sustained in the unlocking period.

**Figure-2.**
Estimation for days taken to record 100-500 cases on caste-group homogeneity

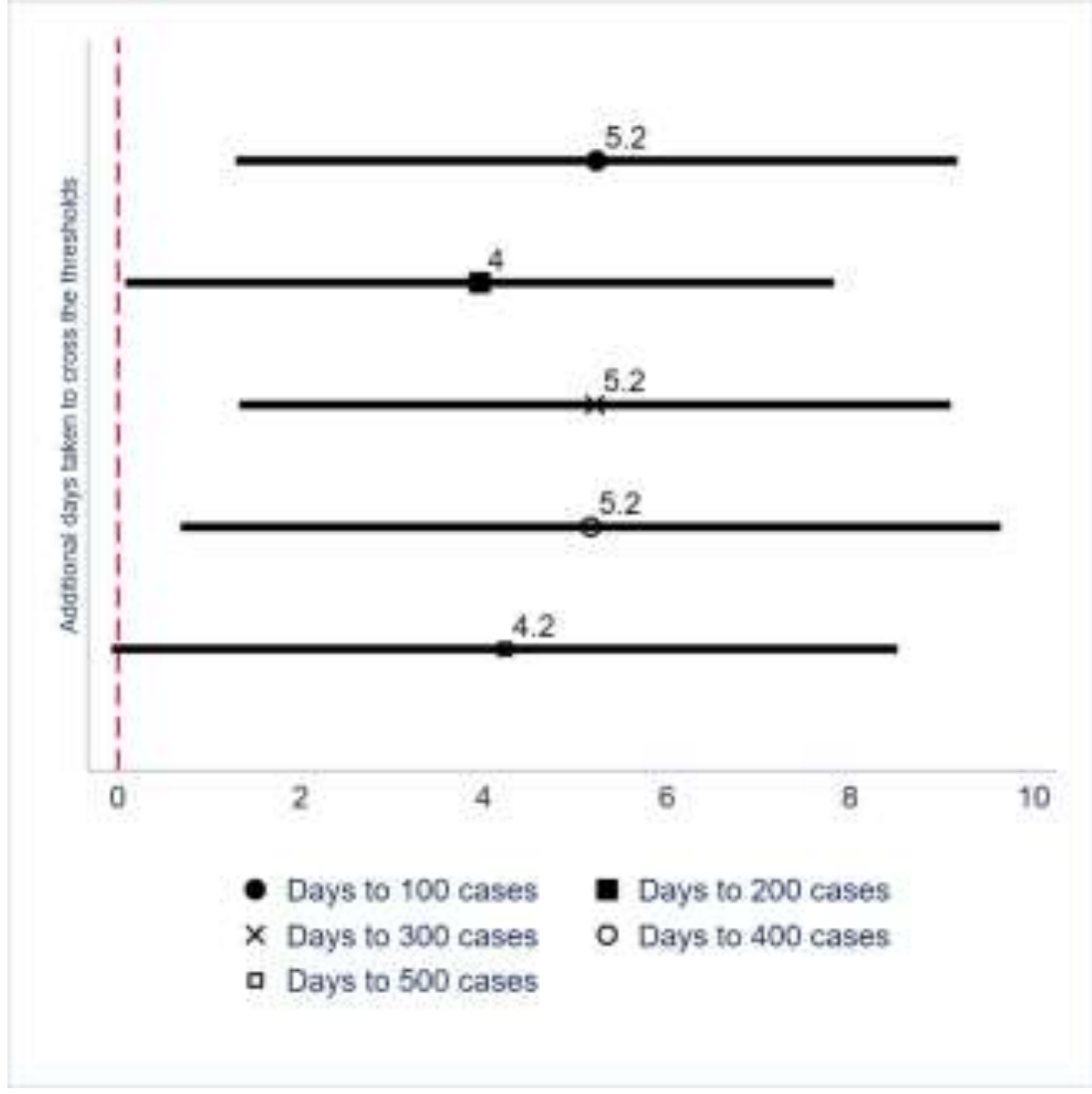

Notes: OLS estimates of the effects of 1 SD increase in caste homogeneity are presented along with 95% confidence intervals calculated from robust standard errors

### *3.2 Intensive Margin*

We now look at the relationship of interest on the intensive margin through daily regression results, as discussed in equation (2). Figure-3 presents the day-wise marginal effects. Here, in addition to controls outlined in the previous specification discussed above, we add lagged total deaths in the past seven days as a share of district-level population and daily fixed effects, which account for time-variant administrative measures and broad COVID-19 diffusion patterns, respectively.[22] The findings reveal that the relationship between caste-group homogeneity and COVID-19 cases per 10,000 population is negative and significant at 1% level for each day of the lockdown over the four phases, ceteris-paribus. In the initial days of the lockdown, the daily effect size was found to be stronger in comparison to those towards the end of the lockdown. On average, during the entire period, the effect ranges from -0.015 to -0.021. Back of the envelope calculations suggests that these effects translate to about 1,744 to 2,452 lower cases per day, on average, for one standard deviation increase in caste-group homogeneity.[23] The statistically significant relationship continues to hold for about eight days after unlocking, albeit with a weaker average effect size of -0.016, which translates to about 1,879 lower cases per day on average and turns insignificant thereafter. The results are statistically similar when the outcome variable (COVID-19 cases based on 7-day rolling averages) is replaced with 3-day rolling averages (Appendix Figure-3).

**Figure-3.**

Relationship between caste-homogeneity and daily COVID-19 cases between 25 March and 7 August 2020

[22] Results with phase-wise introduction of controls are given in Appendix Figure-2.
[23] This is based on 590 districts considered in the analysis assuming an average population of 2 million per district.

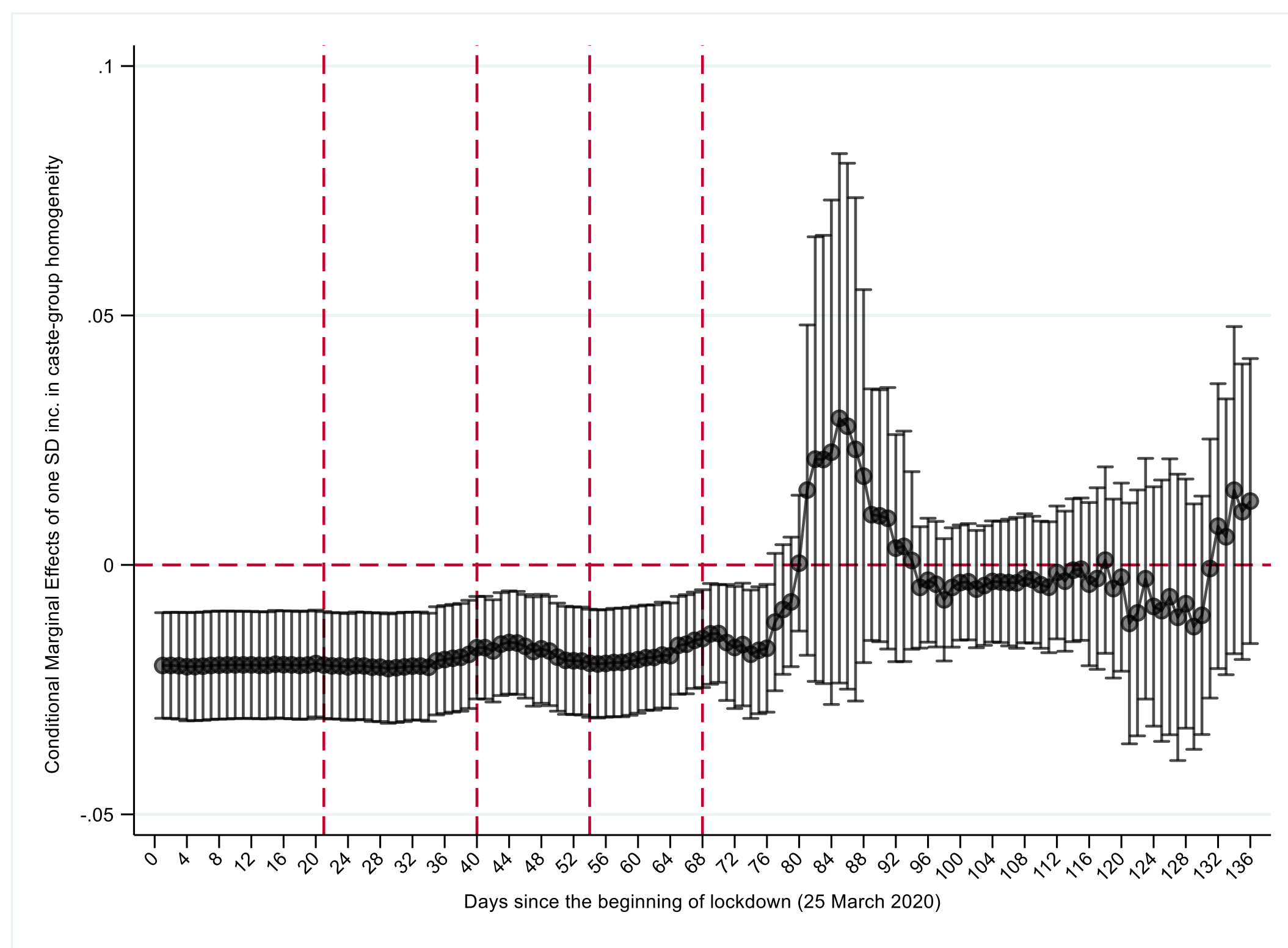

*Note:* Vertical lines denote end of each of the 4 phases of nationwide lockdown in 2020, dependent variable is reported COVID-19 cases per 10,000 population in the district. Marginal effects of caste-group homogeneity*Day are plotted (as in Eq. (2)). 95% confidence intervals calculated after double-clustering of the standard errors by districts (590) and days (!36). All controls outlined in appendix table-1 are used in the regression.

### *3.3 Robustness checks*

#### *3.3.1 State-wise COVID-19 testing*

It is possible that regions having higher caste-group homogeneity may also have systematically lower testing, because of which the reported number of COVID-19 cases in these regions is low. In this case, our findings may be driven by fewer tests in these areas. To examine this possibility, we incorporate daily data on the number of COVID-19 tests conducted across the states during the lockdown and the unlocking periods. Despite missing data for some states in the early period of the lockdown, we are able to use testing data for 32 of 36 states and Union Territories (UT).[24] For missing data, we use a missing dummy that takes the value of 1 for every district whose testing data is not available for a particular day and 0 otherwise. Figure-4 presents the day-wise regression estimates after additionally controlling for the data on COVID-19 testing. The findings remain statistically valid and comparable in terms of effect size to what we observe in Figure-3.

[24] Testing data is fully available for the duration of the analysis for 26 states and UTs and is partially available for six states and UTs.

**Figure-4.**

Relationship between caste-homogeneity and COVID-19 cases between after controlling for state-wise testing data between 25 March and 7 August 2020.

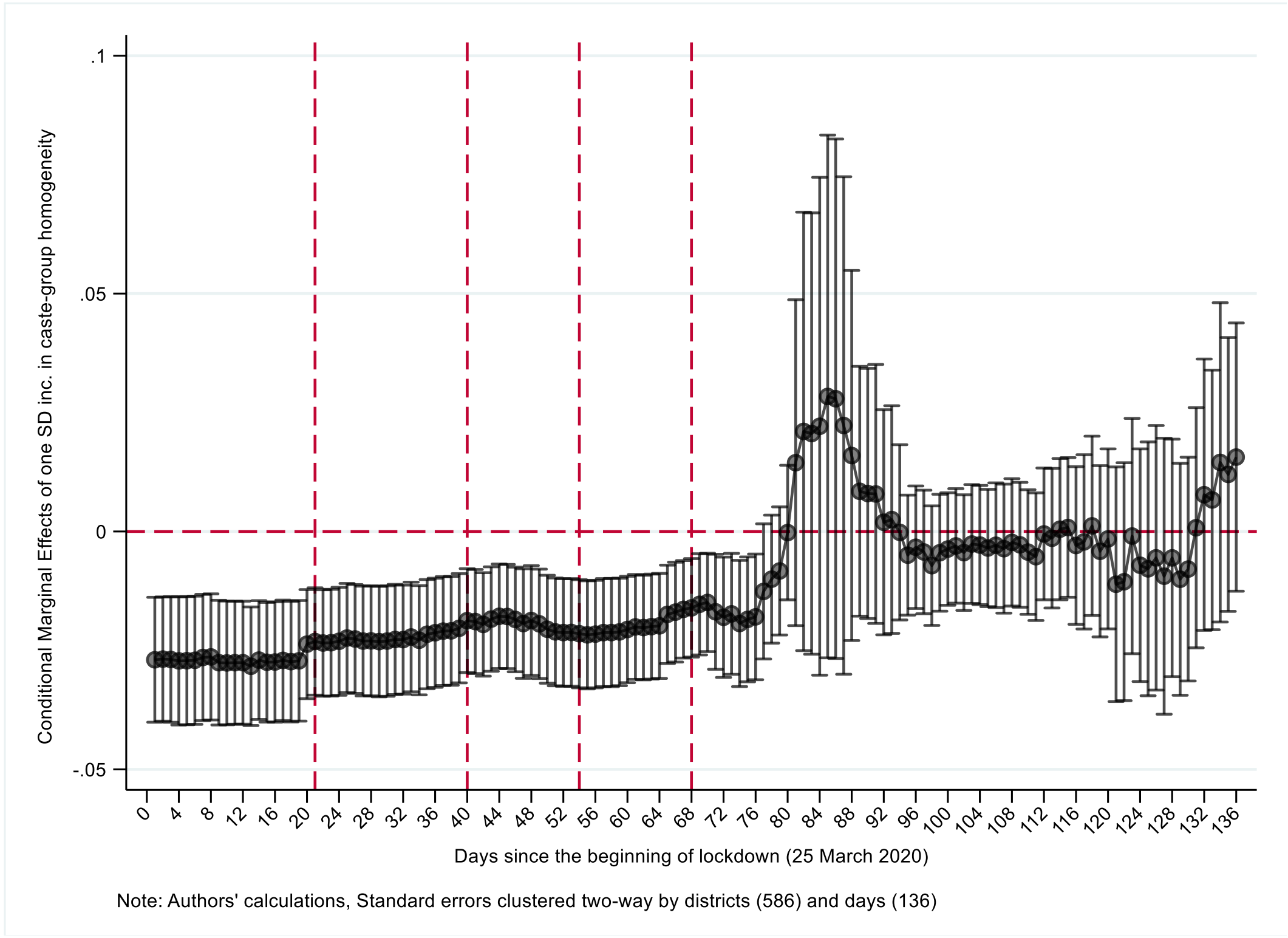

*Note:* Vertical lines denote end of each of the 4 phases of nationwide lockdown in 2020. Dependent variable is reported COVID-19 cases per 10,000 population in the district. Marginal effects of caste-group homogeneity*Day are plotted (as in Eq. (2)). 95% confidence intervals calculated after double-clustering of the standard errors by district and day. All controls outlined in appendix table-1 are used in the regression.

### *3.3.2 Testing for potential unobserved heterogeneities*

To ensure our regression estimates are robust to unobserved heterogeneities emanating through omitted variable bias, we estimate the bias-adjusted treatment effects that use the "selection on observables versus unobservable" as explained earlier. The R-square value obtained from the regression with day interaction (our final specification, as mentioned in equation (2)) is 0.639, which implies that the $R^2_{max}$ will be about 0.830. With this information, we plot the δ obtained for each of the 68 days of the lockdown period (Figure-5) for specification in equation (2). As one can observe, the absolute values of δ range between 2.3 to 2.9 during the lockdown period, which is well above the cut-off value of 1. This indicates that the influence of unobservables will have to be 2.3-2.9 times that from the observables to ensure that the null hypothesis is not rejected. Given the wide range of controls used in the estimation, these exceptionally high values of δ offer credibility to the validity of our estimates. Additionally, we incorporate testing data as done in Figure-4 and re-estimate the δs. The absolute values continue to be well over 1 for each of the 68 days of the lockdown period (1.8 to 2.8).

**Figure-5.**

Accounting for potential OVB (Plotted values of $\delta$)

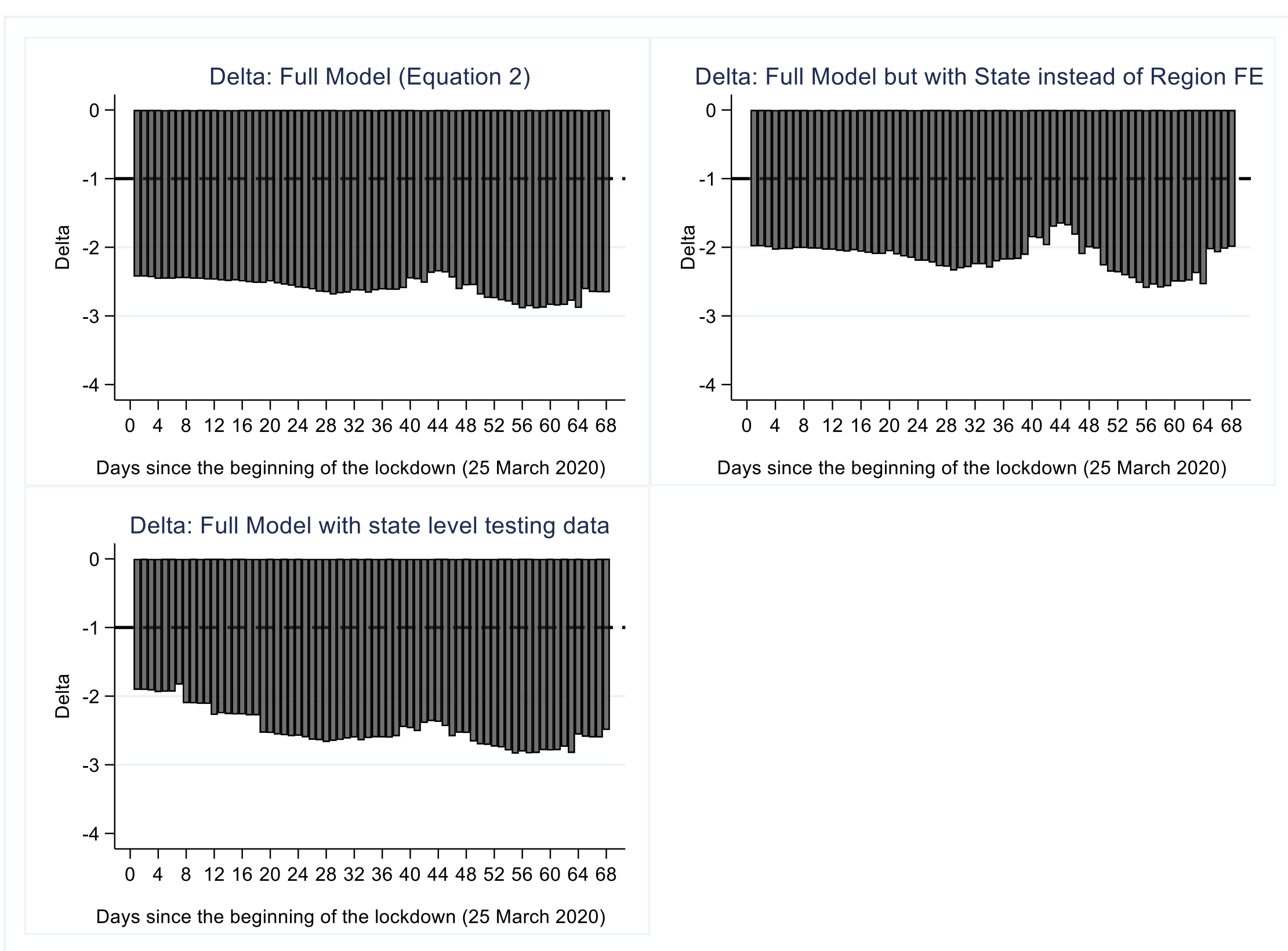

*Note:* "psacalc" command in STATA is used to generate the values of δ. Absolute value of δ = 1 implies that combined effects of unobservables will have to be as large as the effect of all the controls used in the regression put together negate our findings. Even higher absolute values of δ lend further credibility to our estimates as combined influence of unobservables would have to be even larger to invalidate our results.

### *3.3.3 Shannon's entropy measure*

We additionally consider Shannon's entropy measure (SEM) of diversity for district, $d$ instead of $HISG_d$. Mathematically, this can be presented as $SEM_d = \sum_{i=1}^{k} P_i \log_2(1/P_i)$ where $P_i$ is the share of each caste-group in the overall population, which can be divided into $k$ mutually exclusive groups. Unlike the $HISG_d$ that ranges between 0 and 1, with a higher value implying higher homogeneity, $SEM_d$ ranges from 0 to infinity, with higher values representing greater diversity (Frenken et al. 2007). Thus, although $HISG_d$ and $SEM_d$ are inversely related, the coefficients are not directly comparable. The result indicates that the relationship of interest is robust to this measure of diversity as well (Appendix Figure-4). Note that in some of the districts, a few of the caste-groups may not be present, which would make their share in the population zero. As a result, districts for which $SEM_d$ cannot be computed for the 40 districts, which drop out of the analysis. To validate that this selection is not driving our findings, we re-estimate the model using $HISG_d$for the sub-set of 550 districts for which the $SEM_d$ can be computed. We find that the relationship of interest remains robust (Appendix Figure-5).

*3.3.4 Regression with non-capital districts*

State capitals are usually clusters of higher economic activity. As a result, they may witness larger in-flow of migrants from the entire nation, which can also result in higher caste-group diversity in these areas. As a result of higher economic activity and domestic and international movements, these are likely to be the major hotspots of infection spread. Thus, these regions could potentially drive our results. Though we adequately controlled for these characteristics in our specifications by accounting for average distance of the district from the state capital, we now remove these districts (39 out of 590) from our analysis and re-estimate the day-wise regression results for sensitivity check, which is adequately satisfied (Appendix Figure-6).

*3.3.5. Accounting for underreporting of cases*

Emerging evidence from the second wave of infection in India suggests that COVID-19 related deaths in this period (April to July 2021) were systematically underreported (Jha et al., 2022). If such underreporting confounds with caste-group homogeneity, this will introduce bias into our estimates.

Findings of Jha et al. (2022) suggest that on average, predominant share of these underreported deaths (84%) occurred between April and July of 2021, a period outside our study duration. This suggests that underreporting is less of a concern for our estimates. However, for robustness check, we undertake the same analysis using only the states where Jha et al. (2022) estimated excess (unaccounted) death to be comparatively lower.[25] Our results are statistically robust even if we restrict our analysis to these select states (Figure-6).

**Figure-6.**

Estimates for states where underreporting of deaths was comparatively lower

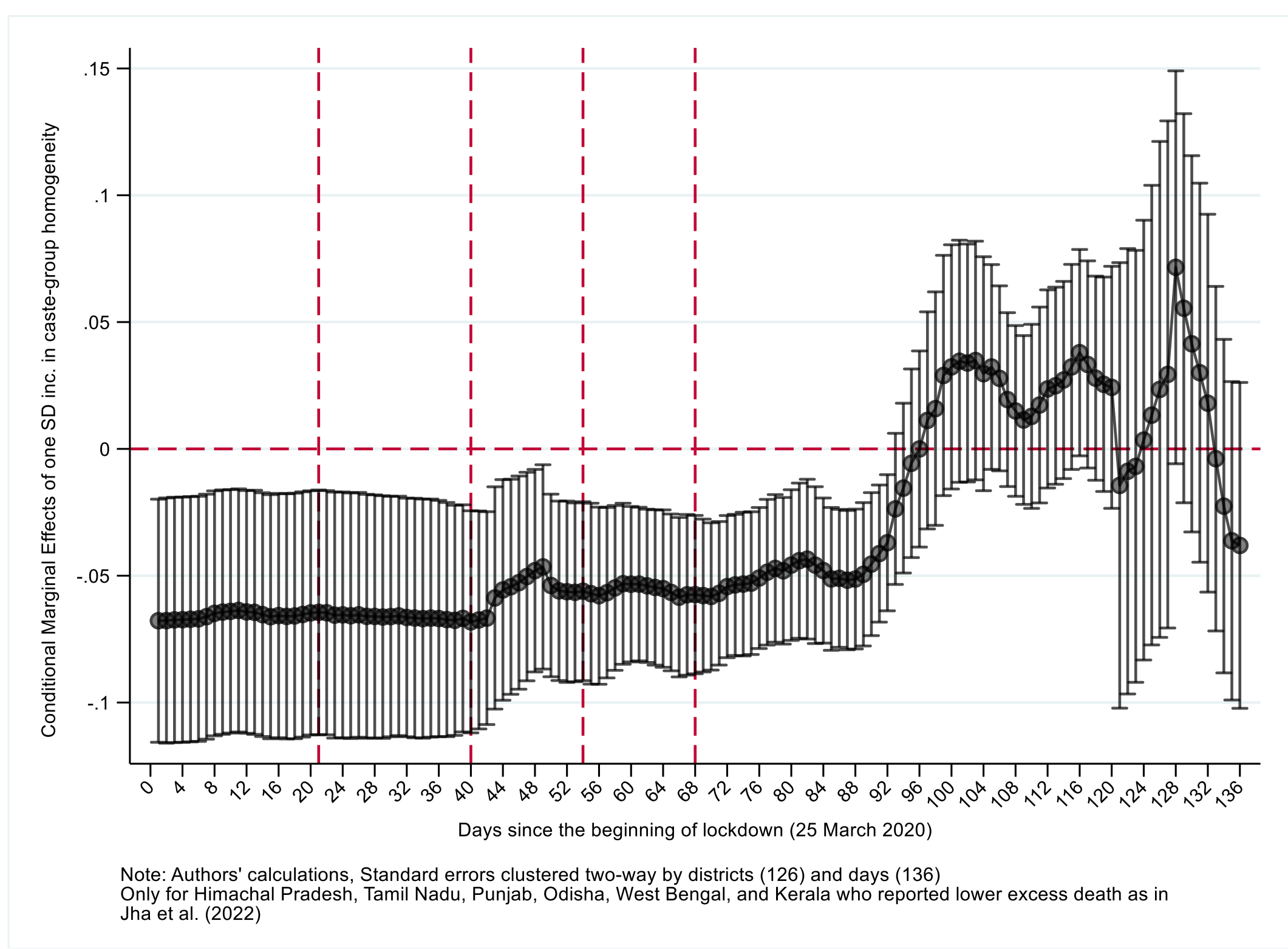

*Note:* The regression sample includes only the states of Himachal Pradesh, Tamil Nadu, Punjab, Odisha, West Bengal and Kerala, where excess death ratio is <=100%. Vertical lines denote end of each of the 4 phases of nationwide lockdown in 2020. Dependent variable is reported COVID-19 cases per 10,000 population in the district. Marginal effects of caste-group homogeneity*Day are plotted (as in Eq. (2)). 95% confidence intervals calculated after double-clustering of the standard errors by district and day. All controls outlined in appendix table 1 are used in the regression.

### *3.4 Exploring Potential Pathways*

[25] This includes states where estimates of excess death was less than 100% (Himachal Pradesh, Tamil Nadu, Punjab, Odisha, West Bengal and Kerala).

As indicated earlier, various population characteristics associated with caste-homogeneous districts could have individually or jointly contributed to the lower spread of COVID-19 cases. We now provide suggestive evidence of pathways that may have contributed to slower infection spread in caste-group homogeneous areas.

### *3.4.1 Engagement with local health workers*

The first channel studies the potential involvement of local health workers in curbing the spread of the infection. In India, the role of frontline health workers has been paramount in running successful vaccination campaigns and eradicating life-threatening diseases like Polio (Deutsch et al. 2017). During the pandemic, ground reports documented that community health care workers played a major role in spreading awareness, communicating health protocols and facilitating treatment for COVID-19 patients.[26]

Simultaneously, there were media reports documenting frontline workers on COVID-19 duty being attacked in some areas, which may have affected their level of motivation and engagement during the pandemic. To check if such instances varied across district-level caste-group homogeneity, we conducted preliminary text analysis and identified multiple news reports from 16 districts and found that 13 of these had median or lower level of caste-group homogeneity (Table-2).[27] This would then suggest that after controlling for human health infrastructure, engagement with and trust on local health workers could have been systematically lower in more caste-group diverse areas and may have mediated differential spread of COVID-19 cases across Indian districts.

**Table-2.** Summary of text analysis on violence against frontline workers across Indian districts

| Sr No | Date | District mentioned | Phrase | Decile of zHHI (Data) |
|---|---|---|---|---|
| **1** | 03-04-2020 | Indore | *"Videos of health care workers being pelted with stones as they are chased away from a locality in India went viral on Thursday, taking the country by storm and garnering global attention. The incident occurred in* | 2 |

[26] Please refer https://www.gavi.org/vaccineswork/iwd2021/international-womens-day-caring-everyone-asha-workers-covid-19-story (accessed on March 24, 2022)
[27] This was undertaken by using a combination of keyword search using 'India', 'Covid' and 'violence' with 'frontline workers' or 'health workers' on google news.

| | | | | |
|---|---|---|---|---|
| | | | *Indore, a city in India's central Madhya Pradesh state, and the area was Tatpatti Bakhal, a hotspot of coronavirus cases in the city."* | |
| **2** | Report 2021 | North East Delhi, South Delhi, Bhopal | *"Several ASHA workers were attacked in South and North-east Delhi; Hajira Baji , an Anganwadi worker Bhopal district Madhya Pradesh, noted that media disinformation linking the spread of Coronavirus with Muslim community after the Tablighi Jamaat incident adversely impacted the Anganwadi workers who were Muslims ."* | 3,3,3 |
| **3** | 12-06-2020 | Bengaluru (urban), Gulbarga (Kalaburagi), Bagalkot, Chikaballapur and Belgaum (Belagavi); Kamrup district | *"ASHAs across Bengaluru (urban), Gulbarga, Bagalkot, Chikaballapur and Belgaum districts of Karnataka were attacked by community members while on COVID-19 duty;*<br><br>*"We are called 'Corona walas' (Corona carriers) and kept at a distance by our neighbours because we are in touch with community members without proper protective gear", said Minara Begum, an ASHA from Kamrup district in Assam"* | 4, 9, 9, 2, 2, 3 |
| **4** | 08-06-2020 | Gorakhpur | *"An Asha worker in Gorakhpur was allegedly abused and assaulted when she had gone to a village to collect details of local residents for COVID-19 vaccination."* | 5 |
| **5** | 22-06-2021 | Thane | *"When Sudarshana Satvi, a 40-year-old Accredited Social Health Activist (ASHA) worker, along with community health officer Prakash Kawade and his team, arrived at Ganje Dhekale, a remote village in Palghar Taluka, in the Thane district of Maharashtra, on June 1, they were met with extreme hostility."* | 4 |
| **6** | 22-04-2020 | Nagpur | *"Health workers visiting some coronavirus containment zones in Maharashtra's Nagpur have complained that they were spat upon by some people, pelted with stones and abused at times while conducting surveys."* | 1 |
| **7** | 11-05-2020 | Faridabad, Gurugram | *"Gurugram On April 23, Rekha Sharma, 39, an ASHA (Accredited Social Health Activists) worker in Ballabgarh, had gone to a society in Sector 65, Faridabad, to conduct a door-to-door survey for coronavirus disease (COVID-19) outbreak. As she was questioning the members of a family about their travel history, they started insulting her. Within minutes, a crowd gathered around her and started abusing her."* | 4,5 |
| **8** | 31-05-2021 | Mahoba | *"Another ASHA worker working in a Mahoba village told Newsclick over the phone that people in her village had been threatening her if she carried out testing or vaccination there"* | 7 |

Note: hyperlinks for specific articles provided above.

This can happen because of multiple reasons. First, as these workers are locally appointed, we expect that in a relatively homogeneous neighborhood with greater interpersonal trust, involvement of community workers and confidence in them would also be higher. Second, when these workers operate in homogeneous areas with similar beliefs, preferences and levels of risk aversion, the marginal effort needed for effective engagement is likely to be lower. Finally, because of potentially greater collective action in homogenous communities, collective demand for health intervention might be higher. Given their historical role in promoting health outcomes, it is possible that caste-group homogeneous regions, which observed stronger frontline worker engagement experienced slower growth of infection during the early period of the outbreak.

To assess this aspect, we utilize the NFHS-5 survey, which collected data from 636,699 households across 707 districts in 2019-21. The survey asks every woman between age 15 to 49 years in these sampled households on whether they have visited an *angandwadi*, Accredited Social Health Activists (ASHA) or other CHWs in the last three months.[28] We code these women who have visited them at least once as 1 and 0 otherwise and run a probit regression on district-level caste homogeneity. In addition, instead of district-level caste-group homogeneity, we run the same regression with the homogeneity measure defined at the Primary Sampling Unit (PSU). The findings in columns 3 and 4 of table-3 indicate a significantly positive association between engagement between local health workers and individuals in caste-group homogenous areas. We repeat the same exercise using NFHS-4 survey data and found similar results (columns 1 and 2 in Table-3), indicating that the above relationship is statistically robust.

Importantly, NFHS-5 also allows us to measure health worker involvement during the pandemic and explore whether this engagement has disproportionately been higher in areas with high caste-group homogeneity. For this, we utilize a crucial characteristic of NFHS-5 survey, which is as follows: In 57 districts from 14 states, some part of survey was conducted between December to

---

[28] *Anganwadi* (hindi) translates to "courtyard shelter". It is childcare center that provide health information, pre-school education, supplementary nutrition and health check-up facilities. ASHA workers are local trained female health activist. For more information, refer https://www.nhm.gov.in/images/pdf/communitisation/task-group-reports/guidelines-on-asha.pdf (accessed on December 18, 2021)

March 2020. The survey was then halted because of the outbreak of the COVID-19 pandemic. After the first wave, the remaining survey resumed in January 2021 and was completed by May 2021. This allows us to make a direct comparison of engagement with community health workers between the sampled households within these 57 districts surveyed before the COVID-19 outbreak with those surveyed post the first wave. By construction, using data only from these 57 districts would account for the time-invariant district-level confounders.

More formally, we use the following regression to estimate the disproportionate changes in engagement with community health workers or ASHAs among those from PSUs with higher caste group homogeneity:

$$Y_{ip} = \alpha + \beta.(Post\ COVIDwave_{ip} * HISG_p) + \gamma Post\ COVID_{ip} + \delta HISG_p + \varepsilon_{ip} \quad (3)$$

Here, $Y_{ip}$ is the outcome variable, which takes the value of 1 if woman, $i$ between age 15 to 49 years from PSU, $p$ has visited an *angandwadi*, ASHA or other community health worker in the last three months prior to the survey and 0 otherwise. $Post\ COVIDwave_{ip}$ is 1 if the survey data of the household falls in between January 2021 to May 2021 (post first wave) and 0 if it falls in between December 2019 to March 2020 (before outbreak). $HISG_p$ is the caste group homogeneity defined at the PSU level. $\varepsilon_{ip}$ is the error term. We are interested in estimating $\beta$, which is the coefficient associated with the interaction of $Post\ COVIDwave_{ip}$ and $HISG_p$.

Table-3 presents the estimation results from above regression (column 5 and 6). We find that the engagement with local health workers has been significantly higher after the start of the pandemic among those from PSUs with higher caste-group homogeneity. Importantly, the results remain similar if we use district-level caste-group homogeneity measures. This provides suggestive evidence of slower COVID-19 spread being mediated via greater engagement of local health workers in caste homogenous areas, which also has been disproportionately higher after the COVID-19 outbreak.

*3.4.2 Higher coordinated action in homogeneous communities*

We also posit that greater coordinated action because of higher cohesion in socially homogenous areas could have led to a slower spread of the virus. If so, socialization and involvement in community-level activities among the residents of caste-group homogeneous regions should also be higher in the pre-pandemic time-period. To examine this, we utilize data from the Time-Use Survey (TUS) conducted by the National Sample Survey Organization of India in 2019.[29] The survey collects information on time allocated to different activities by the members of the sampled households on the day prior to the survey. From the survey, we use two relevant measures of involvement in community-level activities: (i) socializing (socializing and communication, community participation and religious practice) and (ii) unpaid volunteering work (unpaid volunteer, trainee and other unpaid work). As in the earlier case, we consider caste-group homogeneity defined at the PSU and district level: For individuals aged 18-50 years, we run similar regressions of time allocated to socializing and unpaid volunteering work on caste-group homogeneity. We find a positive and significant relationship indicating that individuals from caste-group homogeneous regions, at the local or district level, spent more time socializing and being involved in community-level activities. During the lockdown, it is possible that this social capital and community involvement was instrumental in slowing the spread of the infection. Nevertheless, we are unable to observe this relationship directly and hence we can only provide suggestive evidence on the same.

Table 3: Potential pathways

| | All districts | | | | 57 districts pre-post COVID-19 | | All districts | |
|---|---|---|---|---|---|---|---|---|
| | NFHS 4 (2015-16) | | NFHS 5 (2019-21) | | | | TUS (2019) | |
| | (1) | (2) | (3) | (4) | (5) | (6) | (7) | (8) |
| Standardized HHI (PSU level) | 0.005*** | | 0.004*** | | -0.006** | | 0.003*** | |
| | (0.001) | | (0.001) | | (0.003) | | (0.001) | |
| Standardized HHI (district level) | | 0.001 | | 0.02** | | 0.002 | | 0.005*** |
| | | (0.001) | | (0.001) | | (0.003) | | (0.001) |
| Post COVID-19 first wave | | | | | -0.008 | -0.014*** | | |
| | | | | | (0.005) | (0.009) | | |
| Standardized HHI (PSU level)* Post COVID-19 | | | | | -0.009* | | | |

[29] The survey was administered to 138,805 households having 518,751 individuals. For more information on TUS, refer https://mospi.gov.in/web/mospi/time-use-survey (accessed on December 17, 2021)

| | | | | | | | | |
|---|---|---|---|---|---|---|---|---|
| first wave | | | | | | | | |
| | | | | | (0.005) | | | |
| Standardized HHI (district level)* Post COVID-19 first wave | | | | | | 0.017*** | | |
| | | | | | | (0.005) | | |
| Controls | ✓ | ✓ | ✓ | ✓ | ✓ | ✓ | ✓ | ✓ |
| Observations | 667,074 | 667,074 | 343,449 | 343,449 | 38,932 | 38,932 | 278,332 | 278,332 |

*Note:* All regressions include controls for age, marital status, education, caste dummies, rural/ urban dummy, wealth quantiles/ log of monthly per capita expenditure and state fixed effects. The marginal effects from OLS regressions are presented along with robust standard errors in the parenthesis. For the estimates presented in (5) and (6), state fixed effects are not used. These estimates are given for 57 districts where the survey was conducted before and after the first wave of COVID-19. To generate the estimates in column (1) and column (2), NFHS-4 2015-16 dataset is used, for columns (3), (4), (5) and (6), NFHS-5 2019-21 is used and to generate estimates given in column (7) and column (8), TUS 2019 is used. . *** p-value < 0.01, ** p-value < 0.05, * p-value <0.10.

## 4. Discussion and Policy Implications

The main finding of this paper is that caste-group homogeneity predicts the spread of COVID-19 cases across Indian districts during the nationwide lockdown. Moreover, these gains last for up to a week after the unlocking procedures were initiated. We conduct multiple sensitivity checks to account for various observed and unobserved confounders and find our estimates to be robust. Although the data used does not allow us to gauge the exact underlying mechanism for this relationship, the results are suggestive of the importance of strong community cohesion or "social capital" that may work through caste-group networks and facilitate socially beneficial compliance behavior. In addition, the disproportionately higher involvement of health workers in caste-group homogenous areas after the first wave of the outbreak may have also played a role in curbing the COVID-19 transmission. Unavailability of disaggregated data, particularly community action restricts our ability to assess these underlying mechanisms precisely.

Our analysis covers a period when vaccines were not yet on the horizon and the nature of the pandemic was new to the majority of the population. Hence, compliance with COVID-19 protocols was critical to curtail the spread of infection. Such compliance was likely to be deeply intertwined with community-level cohesion. We thus postulate that social cohesion and reciprocity were critical channels through which these effects were mediated. Importantly, these gains weakened towards the end of the lockdown and became statistically insignificant a week after unlocking began is something that further supports this argument. This is because lockdowns impose economic hardships and fatigue, which increases the cost of sustaining community coordination. This can then reduce socially desired levels of compliance with time. Moreover, easing of lockdown restrictions increased mobility, especially from outside areas, and would have made enforcement and imposing social sanctions more challenging. However, these are conjectures, and a detailed analysis of the underlying pathways remains a question for future research.

Our results have significant policy prescriptions to offer. First, at the onset of such public health crisis, the government should pay extra attention to more vulnerable and socially diverse areas. Our findings provide a basis for identifying and zoning of local areas into socially homogenous and heterogeneous blocks. As a short-term crisis management response, scarce administrative resources (like law enforcement and surveillance) can be prioritized for diverse blocks to compensate for weaker community cohesion, which makes these areas more vulnerable and poses larger spillover risks for the rest of the community. Strengthening the network of community health workers, prioritizing their well-being and incentivizing higher engagement with residents in these diverse areas are some vital interventions that can serve as key policy instruments.

Regarding the administrative response to a sudden public health emergency, identifying areas with weaker community cohesion and allocating scarce resources to these regions is likely to provide a vital window to mount an appropriate response. In this context, local self-governments have a paramount role to play. These local bodies at the rural and urban areas have long been established in countries like India, and they can support decentralized administration under the tutelage of the state and the union governments. These local bodies can help identify the pockets with higher community-level diversity, earmark potential hotspots and take early precautions to avoid more significant outbreaks.

Second, in relatively homogenous areas, as the effect of existing social ties weakens with longer lockdowns and opening up of the economy, strategies to further strengthen community networks might be undertaken. Research indicates that a positive perception about the community is found to enhance private compliance behavior concerning the pandemic (Bicchieri et al. 2021; Das et al. 2021). One possible way to facilitate this could be through decentralized health interventions through community-level representatives that promote greater participation. Here, targeted intervention and educational messages catering to the relevant community through local-level workers may prove to be an important mitigation measure. If implemented, the effectiveness of these policies would be an important area of future research.

Third, we observe that the engagement of CHWs with the households was disproportionately higher in caste-group homogeneous communities after the outbreak, thereby underscoring the key role they could have potentially played in limiting the spread of the infection in these areas. This emphasizes the importance of decentralized health interventions suited to local contexts in formulating effective health policies. To this end, strengthening of institutional capacity primarily through the CWHs assumes significance.

## 5. Conclusion

In a pandemic setting, curtailing the growth of infection needs coordinated community action that can be mediated via social cohesion and strong community ties. However, these are constrained by social diversity that places obstacles in realizing such trust and cooperation. Using daily data on COVID-19 infection at the district level from India, we find that caste-group homogeneity played a significant role in predicting lower growth in cases, especially during the nationwide lockdown imposed from March 25 to May 31, 2020. However, these marginal gains appear to dampen over time and evaporate a week after the initiation of the unlocking process. The results are statistically consistent to a range of robustness and internal validity tests. We find suggestive evidence of engagement with CHWs being more effective in caste-group homogeneous areas. These findings offer potential policy levers to the government and public health officials to identify and prioritize allocation of scarce resources to socially diverse regions that are unable to effectively initiate and sustain coordinated community action. We stress on community-strengthening efforts and argue for decentralized public health interventions during pandemics as these are better suited to address differentiated local requirements instead of a ‘one size fits all’ approach.

**Appendix Figure 1.**

Spread of COVID-19 cases across districts of India for each of the four continuous phases of the lockdown (until 31st May 2020)

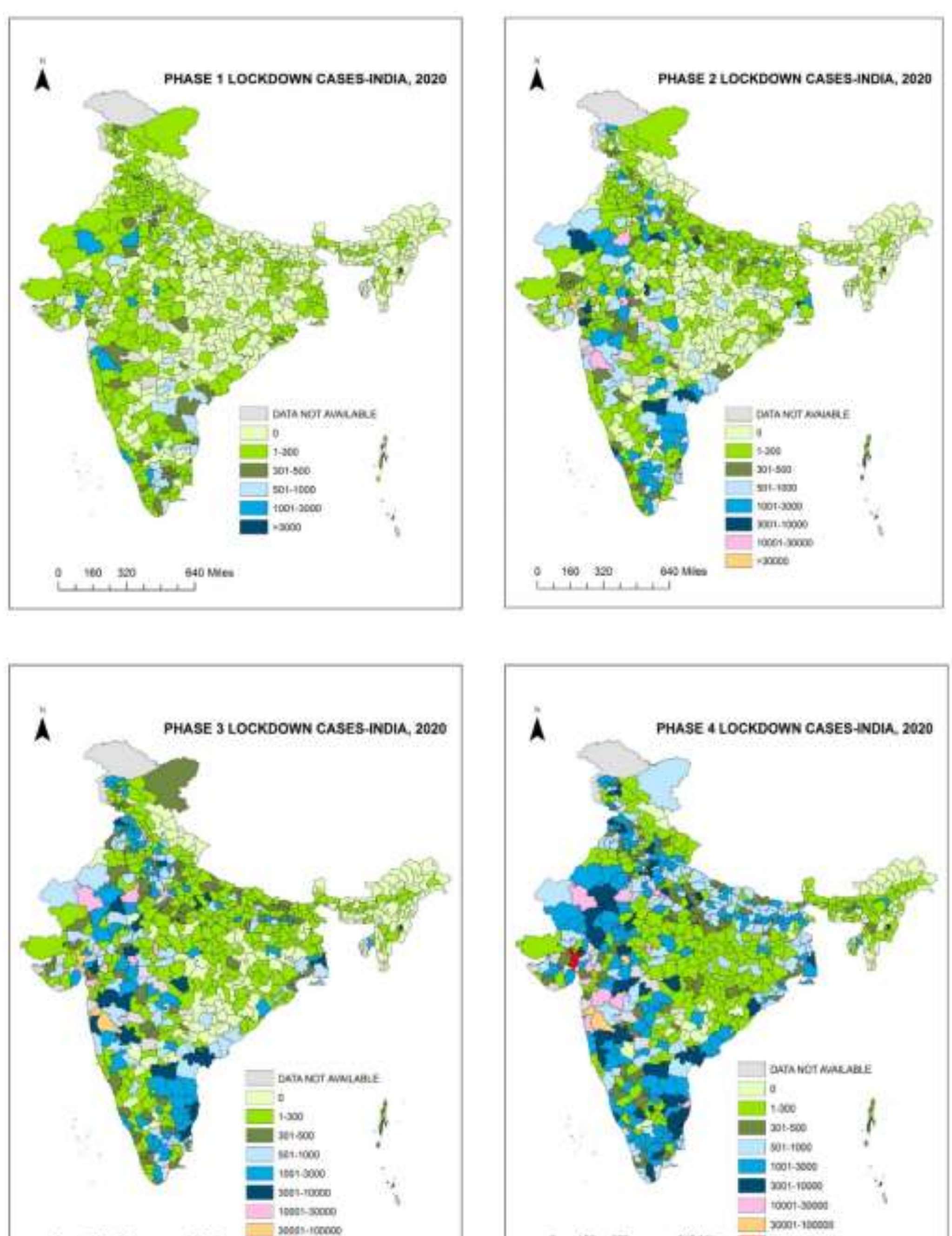

Source: Authors' representation from DDL COVID-19 India database (Asher, et al. 2020). Phase 1: 25 March to April 14; Phase 2: April 15 to May 3; Phase 3: May 4 to May 17 and Phase 4: May 18 till May 31. Numbers based on cumulative cases for each of the phase.

**Appendix Figure 2.**

Relationship between caste-homogeneity and COVID-19 cases between 25 March and 7 August 2020 with phase-wise introduction of controls

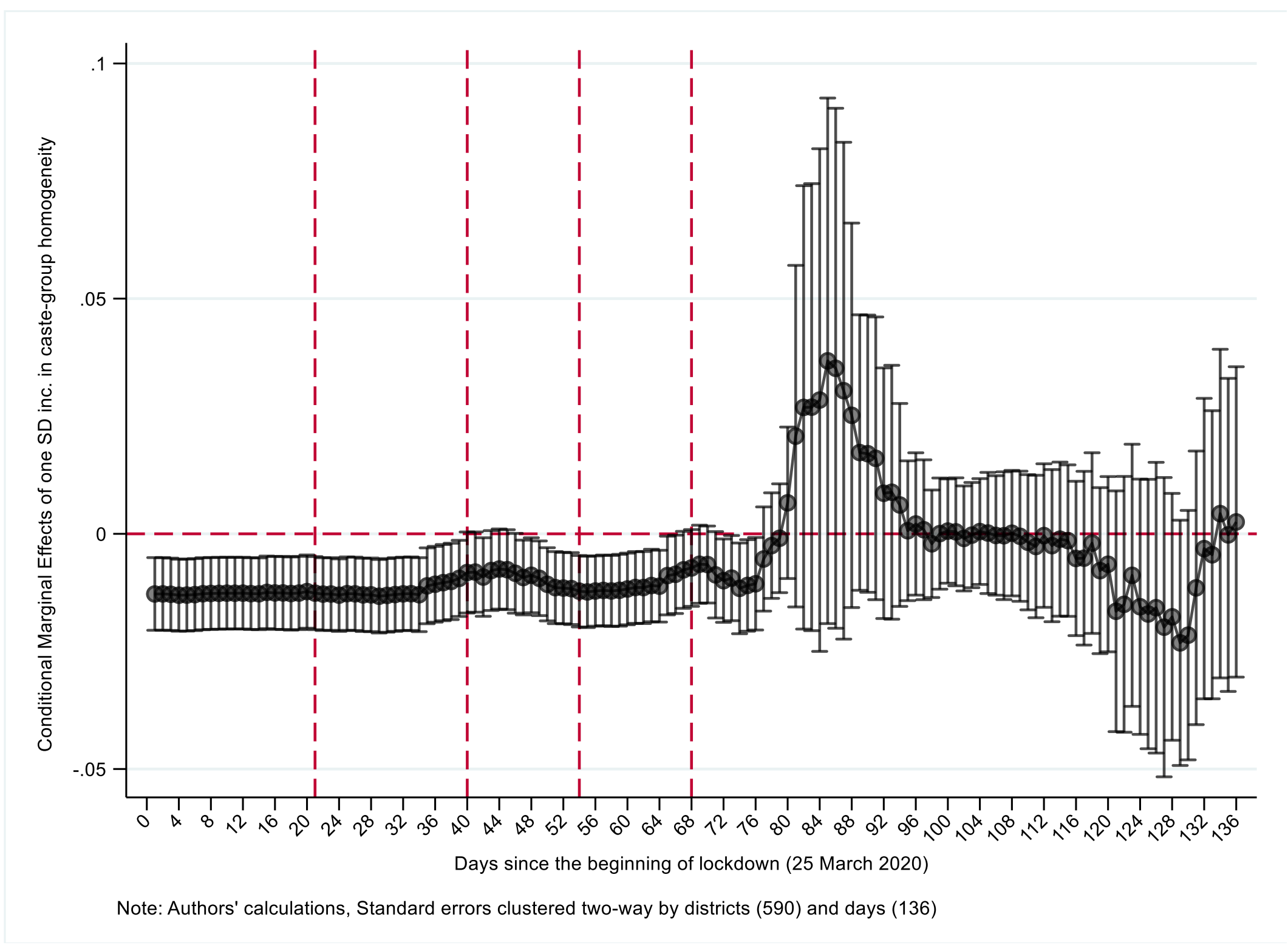

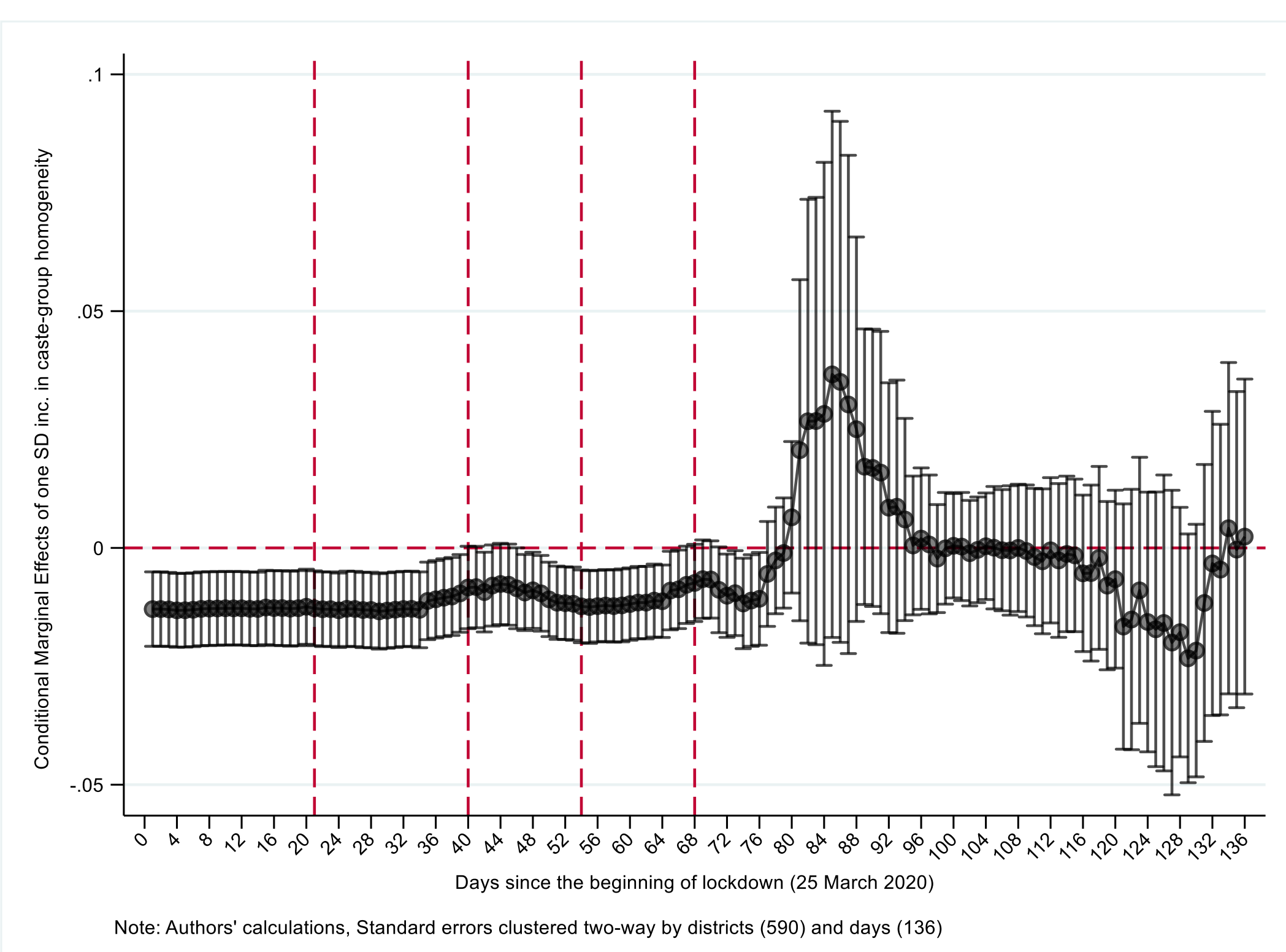

Note: Based on authors' calculations. Vertical lines denote end of each of the 4 phases of nationwide lockdown in 2020, standard errors are clustered two-way across districts (590) and days (136). Model in upper panel controls for: (a) pre-lockdown cases and deaths in past 7 days as a share of population in addition to district level socio-economic controls, (b) demography and health behavior, (c) amenities, migration and mobility controls, respectively. The lower panel additionally controls for physical and human health infrastructure.

**Appendix Figure-3.**

Relationship between caste-homogeneity and COVID-19 cases between 25 March and 7 August 2020 (replacing 7-day rolling cases with 3-day rolling cases)

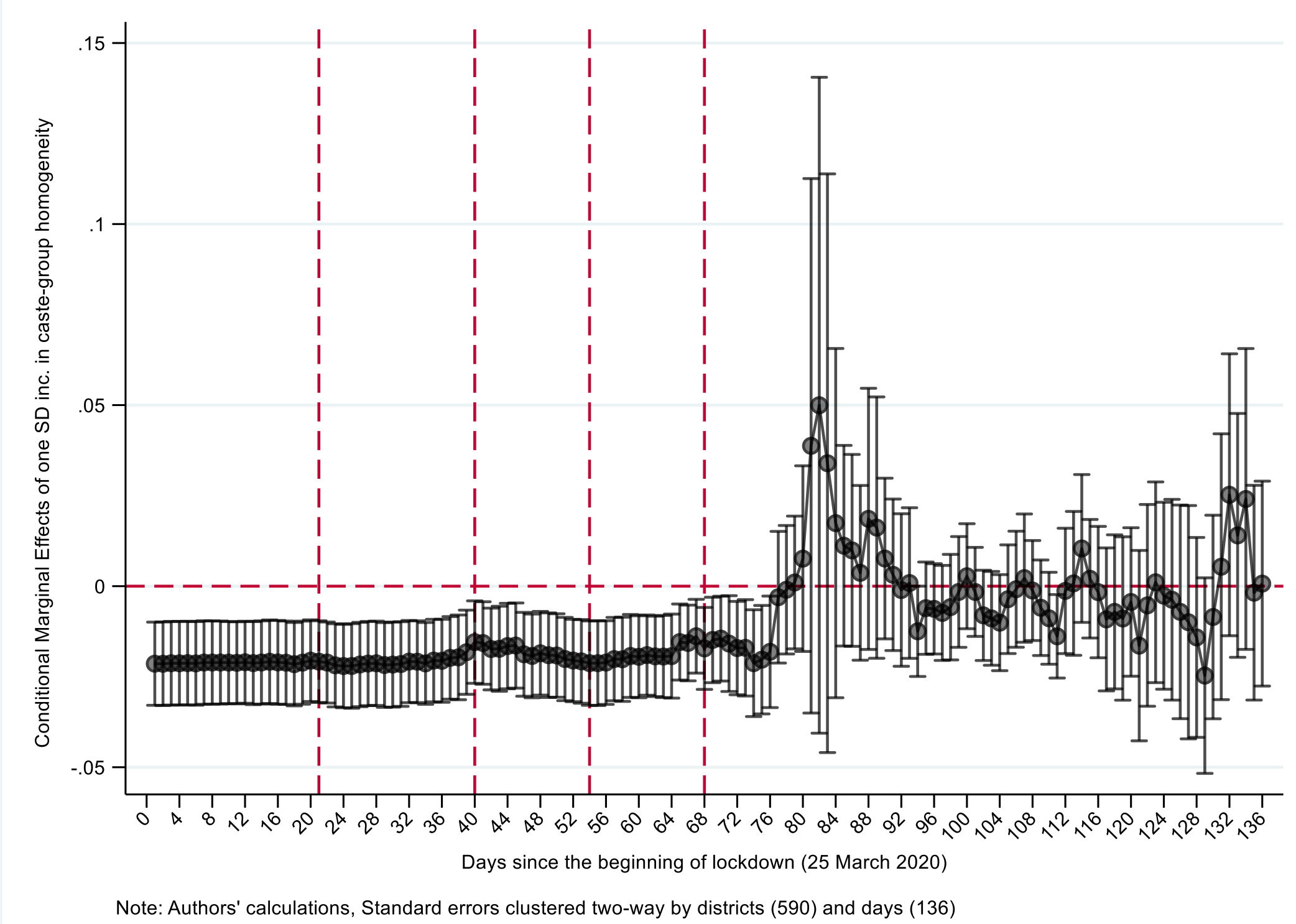

Note: Based on authors' calculations. Vertical lines denote end of each of the 4 phases of nationwide lockdown in 2020.

**Appendix Figure-4.**

Relationship between caste-homogeneity and COVID-19 cases using Shannon's Index of caste-group homogeneity

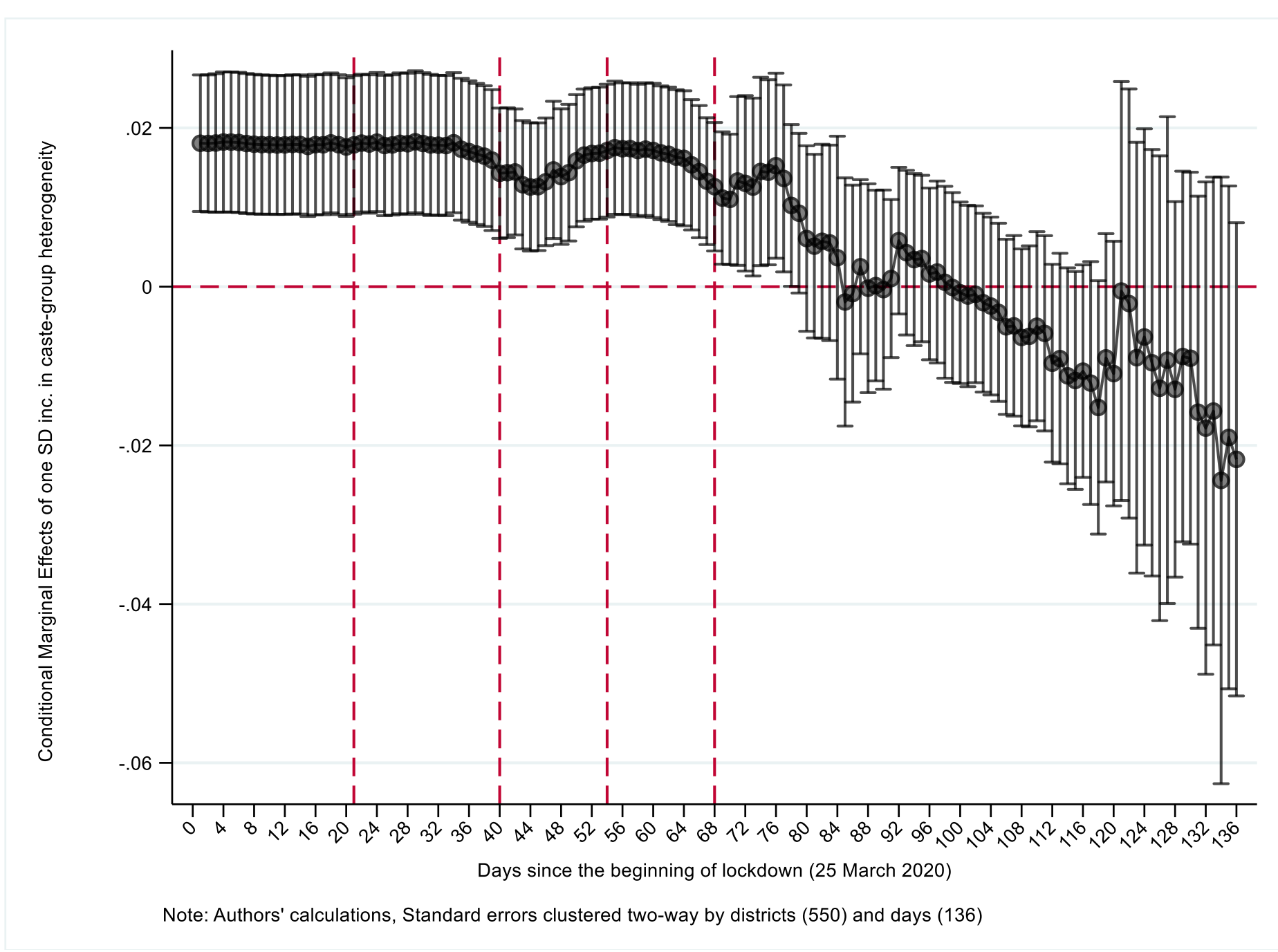

Source: Authors' calculations

Note: Vertical lines denote end of each of the 4 phases of nationwide lockdown in 2020, caste-group heterogeneity computed using Shannon's Index. On the y-axis, the index is now that of heterogeneity and not homogeneity (as in HHI)

**Appendix Figure-5.**

Relationship between caste-homogeneity and COVID-19 cases (HHI computed, and coefficients plotted only for those districts for which Shannon's Entropy Measure for caste-group heterogeneity could be computed)

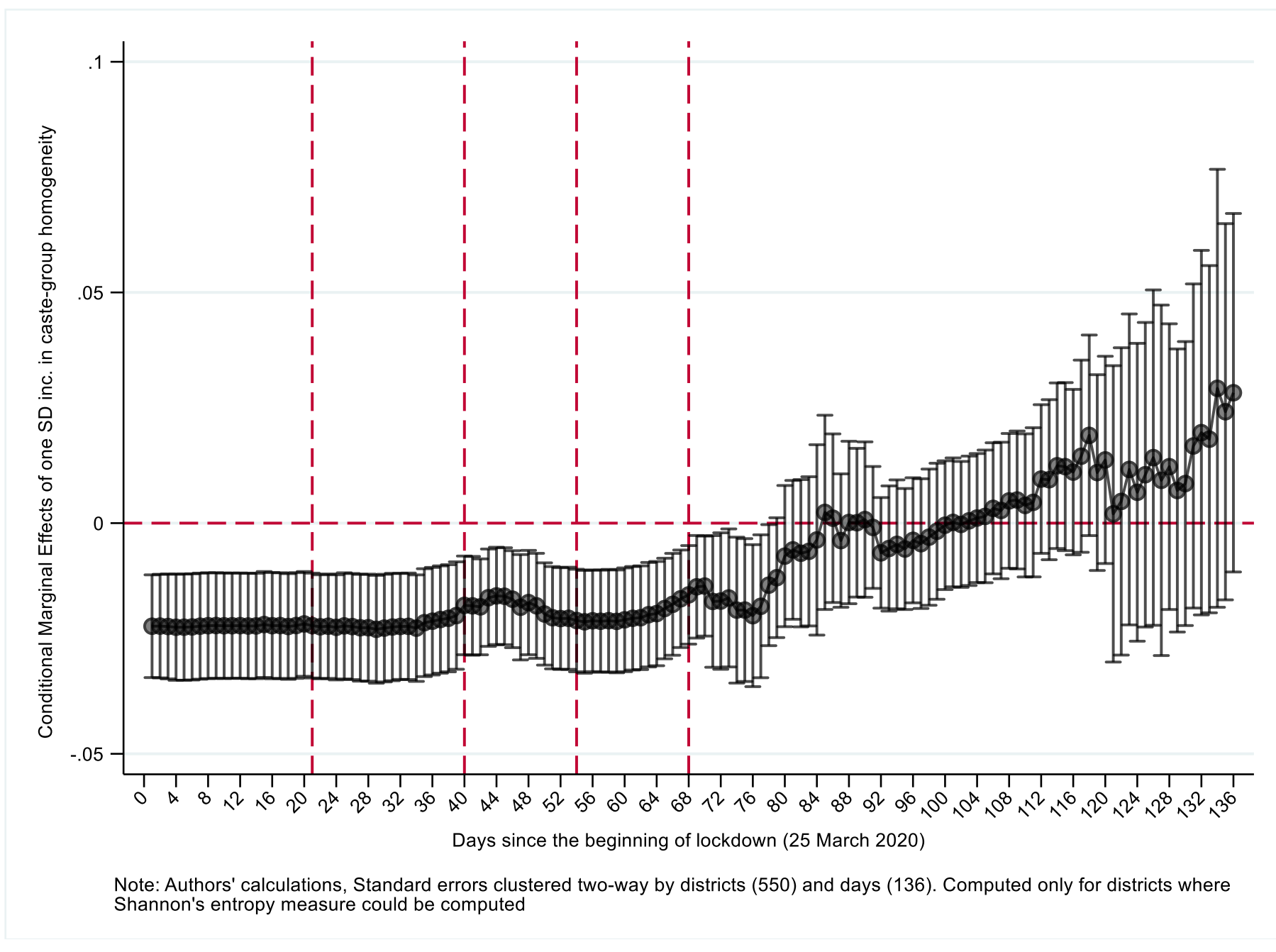

Source: Authors' calculations
Note: Vertical lines denote end of each of the 4 phases of nationwide lockdown in 2020

## Appendix Figure-6.

Relationship between caste-homogeneity and COVID-19 cases between 25 March and 7 August 2020 for non-capital districts

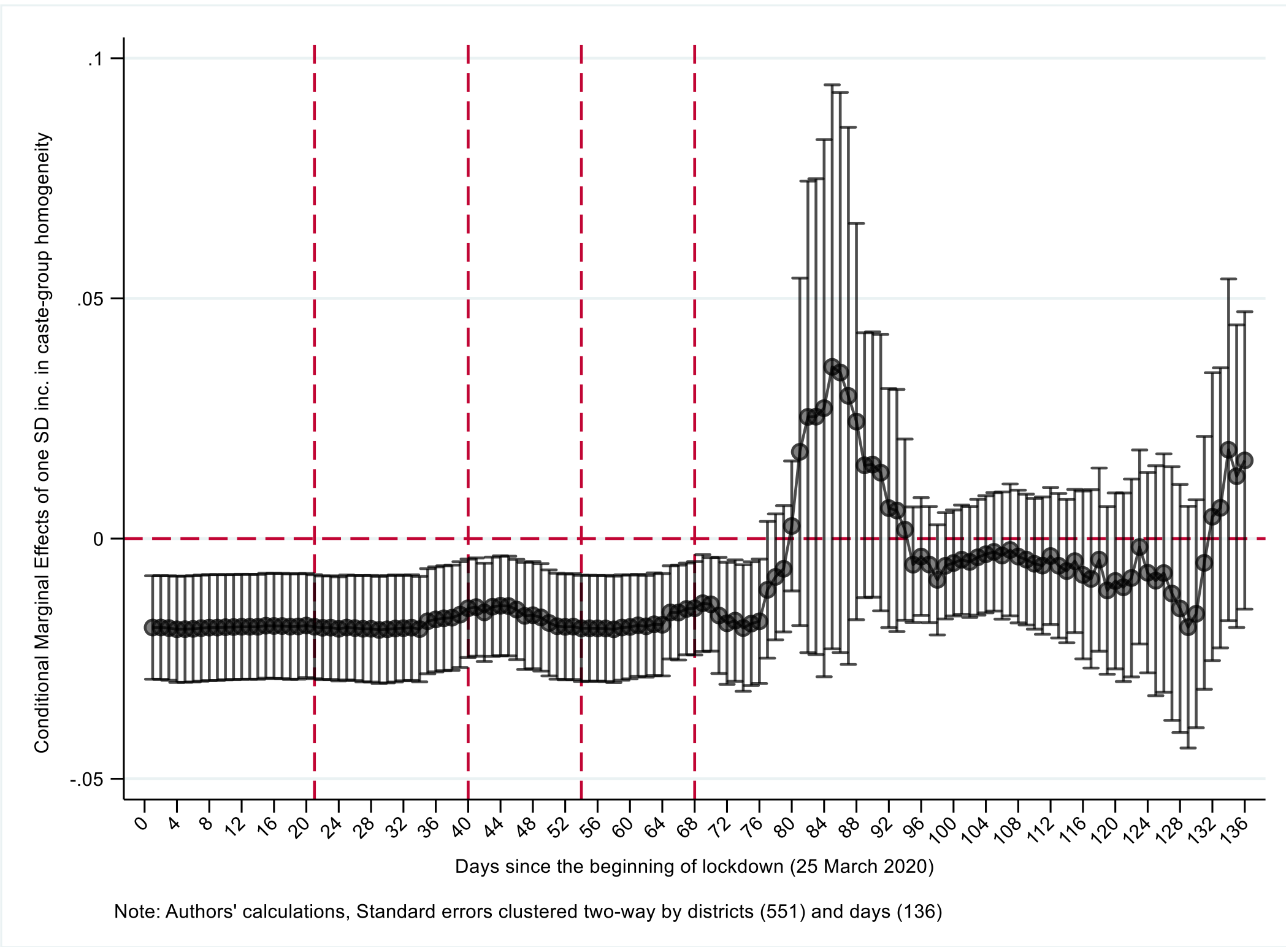

Source: Authors' calculations

Note: Vertical lines denote end of each of the 4 phases of nationwide lockdown in 2020, capital districts from state, union territories and country excluded (49 out of 590)

## Appendix Table 1.

Description of dependent and independent variables used in analysis

**No. of COVID-19 cases and deaths for each district:** Development Data Lab's (DDL) COVID India database (Asher, Tobias, Matsuura, & Novosad, 2020). A crowd-sourced initiative to document district level COVID-19 cases and deaths in India. Cases are identified from news reports and social media posts from credible source. As much data as is available for each case is manually documented in the database, with each entry accompanied by a source url (covindia.com) serving as a citation for of the data point. To account for any systematic reporting errors, three- and seven-day rolling averages also considered for analysis. The outcome variable is taken as a share of district level population.

**Z-score for Herfindahl Hirschman Index (HHI) for caste-groups:** NFHS-4 (2015-2016). HHI computed using distribution of households across the broad caste groupings of Scheduled Tribes, Scheduled Castes, Other Backward Classes and Others. Scale of the index changed to 0 to 100. Z-score computed using mean and standard deviation from all India mean values. Appropriate weights considered. We also compute the same measure using NSSO's 76th round (2018) for robustness check.

**Any COVID-19 cases before the lockdown:** DDL's COVID India database used to identify any cases before the lockdown. This is critical to address the issue of endogeneity as it controls for various characteristics of districts that became epicentres of COVID-19 pandemic in India. For robustness check, we also take number of cases before the lockdown as a share of population in the district.

**Total no. of COVID-19 deaths in the previous seven days as a share of district population:** This variable is included to control for time-variant administrative response to the spread of the pandemic at the district level. Given the administrative structure of India where each district is under the stewardship of a bureaucrat (district collector) that directly reports to the Chief Minister of a state, allocation of scarce administrative resources in a pandemic is likely to be most sensitive to COVID-19 related casualties incurred. Variable is taken as a share of district level population.

**Density of household in a district:** DDL COVID India database sources information on demographics from the Population Census (2011).

**No. of households in a district:** DDL COVID India database sources information on demographics from the Population Census (2011).

**Proportion of households with elderly members:** NFHS-4 (2015-2016). Proportion of households in a district with at-least one member who is at-least 60 years of age.

**Proportion of households in the district that are Hindus:** NFHS-4 (2015-2016). Share of households in a district that are affiliated with Hinduism.

**Proportion of households with higher education:** NFHS-4 (2015-2016). Proportion of households in a district where at-least one member has education achievement of higher secondary (12th standard) or above.

**Proportion of households where women read newspaper daily:** NFHS-4 (2015-2016). Proportion of households in a district where any woman in the household in age group 15-49 years reads any newspaper on an almost daily basis.

**Proportion of households where women listen to the radio weekly:** NFHS-4 (2015-2016). Proportion of households in a district where any woman in the household in age group 15-49 years listen to the radio at-least once a week.

**Proportion of households where women watch television almost daily:** NFHS-4 (2015-2016). Proportion of households in a district where any woman in the household in age group 15-49 years watches television on an almost every day.

**Households with finished floor, roof and walls:** NFHS-4 (2015-2016). Households in a district where the house structure has finished floor, roof and walls. Variable is taken as a score ranging from 0 to 3 with score 1 assigned for each of the three criteria.

**Proportion of households with no toilets:** NFHS-4 (2015-2016). Share of households in a district with no toilets or where members report using open spaces or fields for defecation.

**Average number of rooms for sleeping in the household:** NFHS-4 (2015-2016). Average number of rooms used for sleeping per household in a district.

**Proportion of households having own houses:** NFHS-4 (2015-2016). Share of households in a district where any member of the household owns the house where the family is residing.

**Standard Wealth Index:** NFHS-4 (2015-2016). Average of the Standard Wealth Index computed at the district level using Principal Component Analysis.

**Gini Inequality:** NFHS-4 (2015-2016). Gini Index computed at the district level for asset inequality.

**Proportion of households with health insurance:** NFHS-4 (2015-2016). Share of households in a district where at-least one member has access to health insurance or is covered under some health scheme.

**Proportion of households where at-least one member smokes:** NFHS-4 (2015-2016). Proportion of households in a district where at-least one member in the household smokes at-least on a weekly basis.

**Proportion of households where members engage in washing hands with soap:** NFHS-4 (2015-2016). Proportion of households in a district hand washing area during the survey was found to have availability of water and soap, or detergent.

**Proportion of households residing at the same location (address) for at-least 10 years:** NFHS-4 (2015-2016). Proportion of households in a district where a man reports to be staying at the same location for at-least 10 years.

**Proportion of households where men have stayed outside their home for one month or more at a time:** NFHS-4 (2015-2016). Proportion of households in a district where a male member in the age-group 15-49 years has stayed outside the home for one month or more at a time in the past 12 months.

**Proportion of households where men have stayed outside their home for six months or more at a time:** NFHS-4 (2015-2016). Proportion of households in a district where a male member in the age-group 15-49 years has stayed outside the home for six month or more at a time in the past 12 months.

**Proportion of households with improved drinking water facilities:** NFHS-4 (2015-2016). Proportion of households in a district with improved water source. Improved water source definition comes from WHO. These include piped household water connection located inside the user's dwelling, plot or yard, other improved drinking water sources: Public taps or standpipes, tube wells or boreholes, protected dug wells, protected springs and rainwater collection.

https://www.who.int/water_sanitation_health/monitoring/water.pdf

**Proportion of households with piped water source:** NFHS-4 (2015-2016). Share of households in a district with piped water inside the house or yard.

**Proportion of households with clean cooking fuel:** NFHS-4 (2015-2016). Share of households in a district who use LPG as cooking fuel.

**Proportion of households with access to below poverty line assistance:** NFHS-4 (2015-2016). Share of households in a district who possess a BPL card.

**Proportion of households with Aadhar:** NFHS-4 (2015-2016). Share of households in a district where every member in the household possesses unique identification card (Aadhar). This is a 12-digit identification number issued by the United Identification Authority of India on behalf of the Union Government of India.

**Proportion of households with access to banking:** NFHS-4 (2015-2016). Share of households in a district where any member of the household owns a bank account or a post office (bank) account.

**Proportion of households that visit public health centres when sick:** NFHS-4 (2015-2016). Share of households in a district where members usually go to the public health centres when sick.

**Allopathic hospitals per ten thousand HH in district:** Census (2011). Total allopathic hospitals per ten thousand people in a district.

**Allopathic doctors per million HH in district:** Census (2011). Total allopathic doctors per million people in a district.

**Distance from state capital:** Google Earth Pro. Distance between districts and state capitals in kilometres (in 00s).

**Total flights from the district in the period of analysis:** Airport Authority of India. Total number of flights from the district between April and July 2020 (in 00s).

Note: Appropriate weights considered in district wise calculations from sample surveys.

**Appendix Table 2.**

Summary Statistics

| | Below median caste-group homogeneity (less homogeneous) | | Median & above caste-group homogeneity (more homogeneous) | | Total | | |
|---|---|---|---|---|---|---|---|
| 7-day rolling cases per 10,000 population in the district | Mean | SD | Mean | SD | Mean | SD | Difference |
| Period-1: 25 March - 14 April (2020) | 0.045 | 0.115 | 0.043 | 0.101 | 0.044 | 0.108 | 0.002 |
| Period-2:15 April-3 May (2020) | 0.112 | 0.343 | 0.074 | 0.283 | 0.093 | 0.315 | 0.038 |
| Period-3: 4 May-17 May (2020) | 0.185 | 0.426 | 0.189 | 0.761 | 0.187 | 0.616 | -0.003 |
| Period-4: 18 May- 31 May (2020) | 0.290 | 0.592 | 0.327 | 1.046 | 0.309 | 0.849 | -0.037 |
| Period-5: 1 June - 14 June (2020) | 0.587 | 1.459 | 0.634 | 2.051 | 0.611 | 1.778 | -0.046 |
| Period-6: 15 June 28 June (2020) | 0.767 | 1.730 | 0.935 | 3.496 | 0.851 | 2.757 | -0.167 |
| Period-7: 29 June-12 July (2020) | 1.225 | 2.651 | 1.415 | 3.690 | 1.320 | 3.211 | -0.190 |
| Period-8: 13 July-26 July (2020) | 2.473 | 3.723 | 2.909 | 4.487 | 2.691 | 4.125 | -0.437 |
| Period-9: 27 July-9 August (2020) | 3.386 | 4.565 | 3.958 | 5.464 | 3.672 | 5.038 | -0.572 |
| Summary statistics for controls and variables of interest | Mean | SD | Mean | SD | Mean | SD | Difference |
| Proportion of HH with at-least one member with education level >= higher secondary | 0.239 | 0.090 | 0.248 | 0.107 | 0.243 | 0.099 | -0.009 |
| Proportion of HH where women read newspaper regularly | 0.117 | 0.075 | 0.133 | 0.122 | 0.125 | 0.102 | -0.016$^{*}$ |
| Proportion of HH where women listen to radio every week | 0.139 | 0.100 | 0.175 | 0.134 | 0.157 | 0.120 | -0.037$^{**}$ |
| Proportion of HH where women watch television daily | 0.594 | 0.210 | 0.570 | 0.256 | 0.582 | 0.234 | 0.024 |
| Score of household structure with finished floor, roof and walls (0-3) | 2.086 | 0.517 | 2.121 | 0.499 | 2.103 | 0.508 | -0.035 |
| Average number of rooms used for sleeping | 1.791 | 0.248 | 1.911 | 0.334 | 1.851 | 0.300 | -0.121$^{***}$ |
| Proportion of HH where family living in owned houses | 0.815 | 0.120 | 0.813 | 0.146 | 0.814 | 0.133 | 0.002 |
| Standardized Gini for asset inequality (z-score) | 0.168 | 1.010 | -0.106 | 0.990 | 0.031 | 1.008 | 0.274$^{***}$ |
| Standardized Wealth Index (0-100) | 44.291 | 23.145 | 43.605 | 23.172 | 43.948 | 23.141 | 0.686 |
| Proportion of HH where members covered by health insurance | 0.240 | 0.192 | 0.295 | 0.240 | 0.268 | 0.219 | -0.056$^{***}$ |
| Proportion of Hindu HHs | 0.826 | 0.180 | 0.740 | 0.298 | 0.783 | 0.249 | 0.086$^{***}$ |
| Density of population per square kilometre | 939.085 | 2861.016 | 1066.407 | 3059.378 | 1002.746 | 2960.028 | -127.322 |
| Log of households per district | 12.757 | 0.681 | 12.291 | 1.132 | 12.524 | 0.962 | 0.466$^{***}$ |
| Proportion of HHs with at-least one elderly member (>=60years) | 0.361 | 0.062 | 0.363 | 0.076 | 0.362 | 0.069 | -0.003 |
| Proportion of HHs where members wash hands regularly with soap after defecation | 0.602 | 0.198 | 0.601 | 0.197 | 0.601 | 0.197 | 0.000 |
| Proportion of HHs where some members smoke | 0.443 | 0.159 | 0.456 | 0.177 | 0.450 | 0.168 | -0.013 |
| Proportion of HHs with improved drinking water facilities | 0.880 | 0.113 | 0.883 | 0.121 | 0.882 | 0.117 | -0.003 |
| Proportion of HHs with piped water inside the house | 0.609 | 0.246 | 0.656 | 0.233 | 0.632 | 0.240 | -0.047$^{**}$ |
| Proportion of HHs with no toilets | 0.430 | 0.263 | 0.391 | 0.283 | 0.410 | 0.273 | 0.039$^{*}$ |
| Proportion of HHs using LPG as cooking fuel | 0.362 | 0.223 | 0.369 | 0.228 | 0.365 | 0.225 | -0.007 |
| Proportion of HHs with BPL cards | 0.395 | 0.204 | 0.385 | 0.227 | 0.390 | 0.215 | 0.010 |
| Proportion of HHs where all members have Aadhar card | 0.474 | 0.211 | 0.425 | 0.237 | 0.450 | 0.225 | 0.049$^{***}$ |
| Proportion of HHs with a Bank/ PO account | 0.900 | 0.071 | 0.895 | 0.084 | 0.898 | 0.078 | 0.004 |
| Proportion of HHs where members generally go to public health centres when sick | 0.497 | 0.231 | 0.539 | 0.250 | 0.518 | 0.241 | -0.042$^{**}$ |
| Proportion of HHs where family staying at existing residence for at-least 10 years | 0.875 | 0.093 | 0.847 | 0.105 | 0.861 | 0.100 | 0.027$^{***}$ |
| Proportion of HHs where some men stayed outside their residence for at-least a month in the past year | 0.139 | 0.078 | 0.202 | 0.109 | 0.171 | 0.100 | -0.064$^{***}$ |
| Proportion of HHs where some men stayed outside their residence for at-least six months in the past year | 0.058 | 0.050 | 0.107 | 0.097 | 0.082 | 0.081 | -0.049$^{***}$ |
| Average distance from state capital (in 00 Kms) | 2.655 | 2.859 | 2.470 | 1.832 | 2.563 | 2.401 | 0.184 |
| Total flights landing in the districts between April and July (in 00s) | 1.445 | 7.911 | 0.779 | 5.310 | 1.112 | 6.740 | 0.666 |
| No. of allopathic hospitals per 10,000 HH | 0.538 | 0.517 | 0.661 | 0.918 | 0.599 | 0.747 | -0.123$^{**}$ |
| No. of allopathic doctors per million HH | 329.173 | 331.674 | 481.541 | 597.719 | 405.357 | 488.932 | -152.368$^{***}$ |
| Z-score HHI SG: NFHS-4 | -0.635 | 0.206 | 0.481 | 0.957 | -0.077 | 0.889 | -1.116$^{***}$ |
| Observations | 295 | | 295 | | 590 | | |